\documentclass[a4paper,11pt]{article}
\pdfoutput=1 

\usepackage{jheppub} 

\usepackage{epsfig}
\usepackage{latexsym}
\usepackage{xspace}
\usepackage[utf8]{inputenc}
\usepackage{enumerate}
\usepackage{color}

\usepackage[caption=false,position=top]{subfig}

\usepackage{url}

\newcommand{\eq}[1]{\begin{align} #1 \end{align}}

\newcommand{\dens}[1]{\hat{\rho}_{#1}}
\newcommand{\den}[1]{\hat{\rho}}
\newcommand{\mean}[1]{\langle #1 \rangle}

\newcommand\ddfrac[2]{\frac{\displaystyle #1}{\displaystyle #2}}

\newcommand{\per}[1]{i_{\sigma_{#1}}}

\title{
Cumulants of multiple conserved charges and global conservation laws
}

\author[a]{Volodymyr Vovchenko,}
\affiliation[a]{Nuclear Science Division, Lawrence Berkeley National Laboratory,\\ 1 Cyclotron Road, Berkeley, CA 94720, USA}

\author[b,c]{Roman~V.~Poberezhnyuk,}
\affiliation[b]{Bogolyubov Institute for Theoretical Physics,\\ Metrolohichna  St. 14-b, 03143 Kyiv, Ukraine}
\affiliation[c]{Frankfurt Institute for Advanced Studies, Giersch Science Center,\\ Ruth-Moufang-Str. 1, D-60438 Frankfurt am Main, Germany}

\author[a]{Volker Koch}

\abstract{
We analyze the behavior of cumulants of conserved charges in a subvolume of a thermal system with exact global conservation laws by extending a recently developed subensemble acceptance method~(SAM)~\cite{Vovchenko:2020tsr}
to multiple conserved charges.
Explicit expressions for all diagonal and off-diagonal cumulants up to sixth order that relate them to the grand canonical susceptibilities are obtained. 
The derivation is presented for an arbitrary equation of state with an arbitrary number of different conserved charges.
The global conservation effects cancel out in any ratio of two second order cumulants, in any ratio of two third order cumulants, as well as in a ratio of strongly intensive measures $\Sigma$ and $\Delta$ involving any two conserved charges, making all these quantities particularly suitable for theory-to-experiment comparisons in heavy-ion collisions.
We also show that the same cancellation occurs in correlators of a conserved charge, like the electric charge,
with any non-conserved quantity such as net proton or net kaon number.
The main results of the SAM are illustrated in the framework of the hadron resonance gas model.
We also elucidate how net-proton and net-$\Lambda$ fluctuations are affected by conservation of electric charge and strangeness in addition to baryon number.
}


\begin{document}

\maketitle

\section{Introduction}
\label{sec:intro}

Fluctuations and correlations of conserved charges in statistical systems carry rich information on intrinsic properties of matter. These quantities play a central role in studies of the QCD phase diagram, both in first-principle lattice QCD simulations~\cite{Bazavov:2017dus,Borsanyi:2018grb} and in heavy-ion collision experiments~\cite{Bzdak:2019pkr}.
Event-by-event fluctuations of different quantities are used in the search of the QCD critical point~\cite{Stephanov:1998dy,Stephanov:1999zu,Gazdzicki:2015ska}.
Various correlators of conserved charges, on the other hand, carry information on the relevant QCD
degrees of freedom, such as the baryon-strangeness correlator~\cite{Koch:2005vg}.

Fluctuations and correlations of many different quantities, that include both the conserved charges and various hadron number distributions, have been measured in a number of experiments.
These include measurements of second order cumulants, both diagonal~\cite{Alt:2007jq,Adamczyk:2017wsl,Acharya:2019izy,Adam:2020kzk} and off-diagonal~\cite{Anticic:2013htn,Anticic:2015fla,Adam:2019xmk}, as well as higher-order fluctuation measures~\cite{Adamczyk:2013dal,Adamczyk:2014fia,Adam:2020unf,Adamczewski-Musch:2020slf}.
An important question is how to relate the experimental measurements to theoretical predictions.
For instance, cumulants of the net-proton number cannot be computed in many of the theories, lattice gauge theory in particular, where only the conserved baryon number is accessible.
In such a case one either has to reconstruct net-baryon fluctuations from net-proton measurements~\cite{Kitazawa:2011wh,Kitazawa:2012at}, or directly compare net-proton and net-baryon cumulants, accepting an inevitable systematic error stemming from such an approximation.
Another problem is participant (or volume) fluctuations, which is a source of non-dynamical fluctuations affecting comparisons between theory and experiment~\cite{Gorenstein:2011vq,Skokov:2012ds}.

Perhaps the most important issue is the choice of statistical ensemble.
The vast majority of theories operate in the grand canonical ensemble, where the system can freely exchange conserved charges with a reservoir.
Direct comparison of grand canonical susceptibilities with heavy-ion data is commonplace in the literature~\cite{Karsch:2010ck,Bazavov:2012vg,Borsanyi:2014ewa,Alba:2014eba,Fukushima:2014lfa,Albright:2015uua,Fu:2016tey,Almasi:2017bhq,Vovchenko:2017ayq,Bellwied:2019pxh}.
However, all charges are globally conserved in heavy-ion collisions. 
This would imply that the canonical ensemble is more appropriate than the grand canonical ensemble.
The difference between ensembles does not play a major role if only mean hadron yields are considered in central collisions of heavy ions -- due to the thermodynamic equivalence of statistical ensembles for the averages, the difference between hadron abundances evaluated in different statistical ensembles disappears in large systems.
However, the thermodynamic equivalence of statistical ensembles does not extend to fluctuations, meaning that values of second and higher order cumulants will depend on the choice of the ensemble, no matter how large the system is.

The experimental measurements typically have a limited momentum acceptance, covering only a
fraction of the total momentum space. 
In Ref.~\cite{Koch:2008ia} the necessary conditions to emulate the grand canonical ensemble in heavy-ion collisions have been outlined: measurements should be performed in a rapidity acceptance $\Delta Y_{\rm acc}$ which is, on one hand, large enough to capture all the relevant physics, $\Delta Y_{\rm acc} \gg \Delta Y_{\rm cor}$, where $\Delta Y_{\rm cor}$ characterizes the correlation range in rapidity, while on the other hand, it covers only a small fraction of the whole momentum space such that global conservation laws can be neglected, $\Delta Y_{\rm acc} \ll \Delta Y_{\rm 4\pi}$.
Furthermore, the measurements should cover the entire transverse momentum range.

Global conservation effects are non-negligible whenever $\Delta Y_{\rm acc}$ is comparable to $\Delta Y_{\rm 4\pi}$.
The magnitude of these effects, as well as ways to deal with them, have been studied in the past using a picture of an uncorrelated hadron gas with a single globally conserved charge in a number of papers~\cite{Bleicher:2000ek,Begun:2006uu,Bzdak:2012an,Braun-Munzinger:2016yjz,Rogly:2018kus,Savchuk:2019xfg,Barej:2020ymr,Braun-Munzinger:2020jbk}. 
The analysis in Ref.~\cite{Bzdak:2012an} indicated that the effects of global conservation are sizable already for moderate values of the acceptance fraction $\alpha \equiv \Delta Y_{\rm acc} / \Delta Y_{\rm 4\pi} \lesssim 0.2$, especially for higher-order cumulants.
In our recent work~\cite{Vovchenko:2020tsr}, we introduced a subensemble acceptance method~(SAM) -- a procedure to calculate the cumulants in a presence of a single conserved charge for an arbitrary equation of state.
In Ref.~\cite{Poberezhnyuk:2020ayn} this formalism was applied to fluctuations in vicinity of a critical point.

In the present work, we extend the SAM to equations of state with multiple globally conserved charges, as is appropriate e.g. for QCD with baryon number $B$, electric charge $Q$, and strangeness $S$. In addition to conserved charges, we also explore how cumulants of non-conserved quantities, such as e.g. net-proton number, are affected by multiple global conservation laws.
Within this extended formalism we derive cumulant ratios where effects of global conservation laws are canceled out.
We also show that higher-order measures of a conserved charge distribution, such as kurtosis, are affected by conservation laws involving other conserved charges.

The paper is organized as follows.
Sec.~\ref{sec:subensemble} presents the SAM for multiple conserved charges. 
In Sec.~\ref{sec:HRG} we illustrate the formalism on an example of a hadron resonance gas model.
Discussion and conclusions in Sec.~\ref{sec:concl} close the article.

\section{Formalism}
\label{sec:subensemble}

\subsection{Notation}

We shall use a tensor notation throughout this section. 
Each tensor is denoted by a hat. Where applicable, the number of indices shall determine the tensor rank.
We also adopt the Einstein 
notation, where a repetition of each index implies summation over that index.

Let us have a vector $\hat{Q} = (Q_1,\ldots,Q_N)$ of $N$ independent conserved charges in the system.
Each conserved charge is associated with a chemical potential. The vector of chemical potentials is denoted $\hat{\mu} = (\mu_1,\ldots,\mu_N)$.
In the grand canonical ensemble, GCE, the relation between cumulants $\hat{\kappa}^{\rm gce}$ and susceptibilities $\hat{\chi}$  is straightforward:
\eq{\label{eq:cumudef}
\hat{\chi}_{i_1\ldots i_M}
~=~\frac{\partial^{M}(p/T^4)}{\partial(\mu_{i_1}/T) \, \dots \, \partial(\mu_{i_M}/T)}
~=~\frac{\hat{\kappa}^{\rm gce}_{i_1 \ldots i_M}}{VT^3}~\,, \qquad i_j \in 1 \ldots N.
}
Here $p$ is the pressure, $T$ the temperature and $V$ the volume of the system.
The relation~\eqref{eq:cumudef} applies for an arbitrary cumulant~(susceptibility) of order $M$.
Both the susceptibilities $\hat{\chi}_{i_1 \ldots i_M}$ and the cumulants $\hat{\kappa}_{i_1 \ldots i_M}$ are symmetric with respect to any permutation of their indices.

The notation~\eqref{eq:cumudef} for the susceptibilities is different from the one commonly used in the QCD literature~\cite{Borsanyi:2011sw,Bazavov:2012jq}.
There, the susceptibilities read
\eq{\label{eq:suscdefQCD}
\chi^{Q_1 \ldots Q_N}_{l_1 \ldots l_N} = \frac{\partial^{l_1+\ldots+l_N}(p/T^4)}{\partial(\mu_{Q_1}/T)^{l_1} \, \ldots \, \partial(\mu_{Q_N}/T)^{l_N}}, \qquad l_1+\ldots+l_N = M.
}
The quantities in Eqs.~\eqref{eq:cumudef} and \eqref{eq:suscdefQCD} are equivalent, i.e. $\hat{\chi}_{i_1 \ldots i_M} \equiv \chi^{Q_1 \ldots Q_N}_{l_1 \ldots l_N}$, when the set of indices $(i_1 \ldots i_M)$ contains exactly $l_1$ elements equal to unity, exactly $l_2$ elements equal to two, and so on.
For QCD with three conserved charges, baryon number, electric charge, and strangeness, one has $\hat{Q} = (Q_1,Q_2,Q_3) = (B,Q,S)$.
As an example, we write here a diagonal and an off-diagonal fourth order QCD susceptibilities using the two notations:
\eq{
\chi^B_4 &\equiv \hat{\chi}_{1111}~, \\
\chi^{BQS}_{211} &\equiv \hat{\chi}_{1123}=\hat{\chi}_{1132}=...=\hat{\chi}_{3211}~.
}

\subsection{Subensemble acceptance method}

In the present work we generalize the relation~\eqref{eq:cumudef} to account for the presence of global conservation laws for all of the conserved charges.
Our considerations extend the SAM framework developed in our earlier work~\cite{Vovchenko:2020tsr} to multiple conserved charges.
We would like to mention that the SAM is different from the binomial filter, which has been used in the past to account for global baryon conservation effect in an ideal hadron gas picture~\cite{Bzdak:2012an,Braun-Munzinger:2016yjz}.
While the binomial distribution does provide a useful guidance for understanding how and why the global conservation laws affect the fluctuation measurements, this method is designed to work only in the case of an ideal gas equation of state whereas the SAM is formulated for arbitrary equations of state.
We refer to \cite{Vovchenko:2020tsr} for a more thorough discussion of the differences between the SAM and the binomial acceptance.

Let us 
consider a subvolume $V_1 = \alpha V$ of a uniform thermal system where all the charges are globally conserved.
Following our earlier work~\cite{Vovchenko:2020tsr}, we assume that the subvolume $V_{1}$ as
well as the remaining volume $V_{2}=(1-\alpha)V$ are
both of a macroscopic size, i.e. they are
large compared to correlation length $\xi$, $V_{1} \gg \xi^{3}$ and $V_{2} \gg \xi^{3}$.  
Here $\xi$ refers to any correlation length of relevance to the cumulants of multiple conserved charges under consideration.
As a consequence, one can neglect all interactions at the surface separating the two subsystems, meaning that the total Hamiltonian can be expressed as a sum the subsystem Hamiltonians, i.e. 
\eq{
H = H_1 + H_2 + U_{12} \approx H_1 + H_2~.
}

In this case the canonical ensemble partition
function of the total system with total conserved charge vector $\hat{Q}$  reads~\cite{huang2000statistical}
\eq{\label{eq:Ztot}
Z(T,V,\hat{Q}) = \sum_{\hat{Q}^1} 
Z (T,\alpha V,\hat{Q}^1) \, 
Z (T,\beta V,\hat{Q}-\hat{Q}^1)~.
}
Here $\beta \equiv 1 -\alpha$, and 
the sum goes over all possible values of conserved charges in the first subsystem $\hat{Q}^1$.
The probability $P(\hat{Q}^1)$ to simultaneously find all conserved charges in the subsystem  with volume $V_{1}$ equal to $\hat{Q}^1$ is proportional to the product of the canonical partition functions of the two subsystems:
\eq{\label{eq:PBQS}
P(\hat{Q}^1) &  \propto 
Z(T,\alpha V,\hat{Q}^1) \, 
Z(T,\beta V,\hat{Q}-\hat{Q}^1)~. 
}

In the thermodynamic limit, $V \to \infty$, we have
\eq{
Z(T,V,\hat{Q}) = \exp\left[-\frac{V}{T} \, f(T,\dens{}) \right],
}
where $\dens{} = \hat{Q}/V$ is the vector of densities of all the conserved charges and $f(T,\dens{})$ is the free energy density. 

To evaluate the cumulants $\hat{\kappa}_{i_1 \ldots i_M}[\hat{Q}^1]$ of the distribution of conserved charges $\hat{Q}^1$ inside the subvolume $V_1$ we introduce the cumulant generating function $G_{\hat{Q}^1}(\hat{t})$:
\eq{
G_{\hat{Q}^1}(\hat{t}) & \equiv \ln \mean{e^{\hat{t}_i \hat{Q}^1_i}}  
 =  \ln \left\{ \sum_{\hat{Q}^1} \exp(\hat{t}_i \hat{Q}^1_i) P(\hat{Q}^1) \right\} \nonumber \\ 
& = \ln \left\{ \sum_{\hat{Q}^1} \, e^{\hat{t}_i \hat{Q}^1_i} \, \exp\left[-\frac{\alpha V}{T} \, f(T,\dens{}^1) \right] \exp\left[-\frac{\beta V}{T} \, f(T,\dens{}^2) \right]  \right\} + \tilde{C}~.
}

The cumulants, $\hat{\kappa}_{i_1,\ldots,i_M}[\hat{Q}^1]$, correspond to the Taylor coefficients of $G_{\hat{Q}^1}(\hat{t})$:
\eq{
\hat{\kappa}_{i_1 \ldots i_M}[\hat{Q}^1] = \left. \frac{\partial^{M} G_{\hat{Q}^1}(\hat{t})}{\partial\hat{t}_{i_1} \, \dots \, \partial\hat{t}_{i_M}} \right|_{\hat{t}=0} \equiv \left.  \hat{\tilde{\kappa}}_{i_1,\ldots,i_m}(\hat{t}) \right|_{\hat{t}=0}.
}
Here we introduced a shorthand $\hat{\tilde{\kappa}}_{i_1,\ldots,i_m}(\hat{t})$ for generalized $\hat{t}$-dependent cumulants.

All second and higher-order cumulants can be obtained by differentiating the first order cumulants, $\hat{\tilde{\kappa}}_{i_1}(\hat{t})$, with respect to the components of $\hat{t}$.
The first order cumulants read
\eq{
\hat{\tilde{\kappa}}_{i}(\hat{t}) = \frac{\partial G_{\hat{Q}^1}(\hat{t})}{\partial \hat{t}_{i}}=  \frac{\sum_{\hat{Q}^1} \, \hat{Q}^1_{i} \, \tilde{P}(\hat{Q}^1;\hat{t})}{\sum_{\hat{Q}^1} \, \tilde{P}(\hat{Q}^1;\hat{t})} = \mean{\hat{Q}^1_{i}(\hat{t})}~,
}
with the (un-normalized) $\hat{t}$-dependent $\hat{Q}_1$ probability function,
\eq{
\tilde{P}(\hat{Q}^1;\hat{t}) = \exp\left\{\hat{t}_i \hat{Q}^1_i - V \, \frac{\alpha f(T,\dens{}^1) + \beta f(T,\dens{}^2)}{T}\right\}~.
}

In the thermodynamic limit, the probability $\tilde{P}(\hat{Q}^1;\hat{t})$ is highly peaked at the mean values of $\hat{Q}^1$.
The vector $\mean{\hat{Q}^1_{i}(\hat{t})}$ of the mean values of $\hat{Q}_1$ is, therefore, determined by requiring that all partial derivatives of $\tilde{P}(\hat{Q}^1;\hat{t})$ vanish,
$
\partial \tilde{P} / \partial \hat{Q}^1_{i}
= 0$:
\eq{
\label{eq:t}
\hat{t}_i = \hat{\bar{\mu}}_i[T,\dens{}^1(\hat{t})] - \hat{\bar{\mu}}_i[T,\dens{}^2(\hat{t})], \qquad i = 1,\ldots,N.
}
Here 
\eq{
\hat{\bar{\mu}}_i \equiv \frac{\hat{\mu}_i}{T} = \frac{1}{T} \frac{\partial f}{\partial \hat{\rho}_i}
}
is the reduced chemical potential as a function of temperature and densities of conserved charges,
and $\dens{}^1(\hat{t}) = \mean{\hat{Q}^1(\hat{t})} / (\alpha V)$ and $\dens{}^2(\hat{t}) = \mean{\hat{Q}^2(\hat{t})} / (\beta V) = (\hat{Q} - \mean{\hat{Q}^1(\hat{t})}) / (\beta V)$.
For $\hat{t} = 0$ the solution is:
\eq{
\dens{}^1 = \dens{}^2 = \dens{} = \hat{Q}/V,
}
i.e. the charges are distributed uniformly between the subsystems, as should be the case by construction. 
Given that the first order cumulant of the $i$-th charge is $\hat{\kappa}_{i}[\hat{Q}^1] \equiv \mean{\hat{Q}^1_i} = \alpha V \, \hat{\rho}_i$ and the first order susceptibility equals $\hat{\chi}_i \equiv \hat{\rho}_i / T^3$ by definition, we get
\eq{\label{eq:kappa1final1}
\hat{\kappa}_{i}[\hat{Q}^1] = \alpha \, V \, T^3 \, \hat{\chi}_{i}~, \qquad i = 1,\ldots,N.
}

\subsection{Second order cumulants}

Given the generalized first order cumulant $\hat{\tilde{\kappa}}_{i}$, the second order cumulants
$\hat{\tilde{\kappa}}_{i j}(\hat{t})$ are 
\eq{
\hat{\tilde{\kappa}}_{i j}(\hat{t}) = \frac{\partial \hat{\tilde{\kappa}}_{i}(\hat{t})}{\partial \hat{t}_{j}} \equiv \frac{\partial \mean{\hat{Q}^1_{i}(\hat{t})}}{\partial \hat{t}_{j}}, \qquad i,j = 1,\ldots,N.
}
To evaluate $\hat{\tilde{\kappa}}_{i j}(\hat{t})$ we differentiate Eq.~\eqref{eq:t} with respect to $\hat{t}_j$.
We obtain
\eq{\label{eq:delt1}
\hat{\delta}_{ij} = \frac{\partial \hat{\bar{\mu}}^1_i}{\partial \den{}^1_{j_1}} \,  \frac{\partial \den{}^1_{j_1}}{\partial \mean{\hat{Q}^1_{j_2}}} \,  \frac{\partial \mean{\hat{Q}^1_{j_2} (\hat{t})}}{\partial \hat{t}_{j}}
-
\frac{\partial \hat{\bar{\mu}}^2_i}{\partial \den{}^2_{j_1}} \,  \frac{\partial \den{}^2_{j_1}}{\partial \mean{\hat{Q}^2_{j_2}}} \,  \frac{\partial \mean{\hat{Q}^2_{j_2}}}{\partial \mean{\hat{Q}^1_{j_3}}} \,  \frac{\partial \mean{\hat{Q}^1_{j_3} (\hat{t})}}{\partial \hat{t}_{j}},
}
where we used the chain rule multiple times.
Here one observes that
\eq{
\label{eq:drho1dQ1}
\frac{\partial \den{}^1_{j_1}}{\partial \mean{\hat{Q}^1_{j_2}}}  = 
\frac{\hat{\delta}_{j_1 j_2}}{\alpha V}~,~~~~~~~~~~ 
\frac{\partial \den{}^2_{j_1}}{\partial \mean{\hat{Q}^2_{j_2}}}  = 
\frac{\hat{\delta}_{j_1 j_2}}{\beta V}~,~~~~~~~~~~ 
\frac{\partial \mean{\hat{Q}^2_{j_2}}}{\partial \mean{\hat{Q}^1_{j_3}}}  =
-\hat{\delta}_{j_2 j_3}~,
}
and, by definition,
\eq{\label{eq:dQ1dt}
\frac{\partial \mean{\hat{Q}^1_{j_2} (\hat{t})}}{\partial \hat{t}_{j}} & =
\hat{\tilde{\kappa}}_{j_2 j}(\hat{t})~.
}
Equation~\eqref{eq:delt1}, therefore, reduces to
\eq{\label{eq:kappa2eq}
\hat{\delta}_{ij} = \frac{1}{V} \left[\frac{1}{\alpha} \frac{\partial \hat{\bar{\mu}}^1_i}{\partial \den{}^1_{j_1}} + \frac{1}{\beta} \frac{\partial \hat{\bar{\mu}}^2_i}{\partial \den{}^2_{j_1}}  \right] \, \hat{\tilde{\kappa}}_{j_1 j}(\hat{t})~.
}
The derivatives $\partial \hat{\bar{\mu}}^{1,2}_i / \partial \den{}^{1,2}_{j_1}$ can be expressed in terms of second order susceptibilities, $\hat{\chi}_{ij}$. 
Indeed, since the second order susceptibilities are defined by the derivatives of the conserved number densities,
\eq{\label{eq:chi2def}
\hat{\chi}_{ij} \equiv T^{-3} \, \frac{\partial \den{}_i}{\partial \hat{\bar{\mu}}_j}~,
}
the inverse derivatives, $\partial \hat{\bar{\mu}}_i / \partial \den{}_{j}$, are defined by the inverse matrix of second order susceptibilities,
\eq{\label{eq:dmudrho}
\frac{\partial \hat{\bar{\mu}}_i}{\partial \den{}_j} = T^{-3} \, \hat{\chi}^{-1}_{ij}~.
}
The matrix equation~\eqref{eq:kappa2eq} for the second order cumulants, therefore, reads
\eq{\label{eq:delt3}
\hat{\delta}_{ij} = \frac{1}{VT^3} \left[\frac{1}{\alpha} \hat{\tilde{\chi}}^{'-1}_{ij_1} + \frac{1}{\beta} \hat{\tilde{\chi}}^{''-1}_{ij_1}  \right] \, \hat{\tilde{\kappa}}_{j_1 j}(\hat{t})~.
}
Here 
$\hat{\tilde{\chi}}_{ij_1}^{'}$ and 
$\hat{\tilde{\chi}}_{ij_1}^{''}$ correspond to the matrix of second order conserved charge susceptibilities evaluated at an arbitrary finite $\hat{t}$ in the first and second subsystems, respectively.

The solution to Eq.~\eqref{eq:delt3} is
\eq{\label{eq:kappa2t}
\hat{\tilde{\kappa}}_{ij}(\hat{t}) = V T^3 \, \left[\frac{1}{\alpha} \hat{\tilde{\chi}}^{'-1}_2 + \frac{1}{\beta} \hat{\tilde{\chi}}^{''-1}_{2} \right]^{-1}_{ij}, \qquad i,j = 1,\ldots,N.
}

For $\hat{t} = 0$ one has $\hat{\tilde{\chi}}^{'}_2 = \hat{\tilde{\chi}}^{''}_2 = \hat{\chi}_2 \equiv \hat{\chi}_{ij}$, therefore
\eq{\label{eq:kappa2final1}
\hat{\kappa}_{ij}[\hat{Q}^1] = \alpha V T^3 \, \beta \, \hat{\chi}_{ij}~, \qquad i,j = 1,\ldots,N.
}
We note, that the correction factors due to global charge conservation are the same for all second
order susceptibilities. 
An important consequence of this result is that a ratio of any two second order cumulants of conserved charges coincides with the corresponding ratio of grand canonical susceptibilities, i.e. effects of global conservation are canceled out.

\subsection{Third order cumulants}
\label{sec:kappa3}

The third order cumulants are calculated by differentiating Eq.~\eqref{eq:kappa2t} with respect to $\hat{t}_k$:
\eq{\label{eq:skewdef}
\hat{\tilde{\kappa}}_{i j k}(\hat{t}) = \frac{\partial \hat{\tilde{\kappa}}_{ij}(\hat{t})}{\partial \hat{t}_{k}}~, \qquad i,j,k = 1,\ldots,N.
}

To evaluate this derivative we will make use of the following identity from the matrix calculus:
\eq{\label{eq:invder}
\frac{\partial \hat{U}^{-1}}{ \partial x} = -\hat{U}^{-1} \, \frac{\partial \hat{U}}{\partial x} \, \hat{U}^{-1}.
}
Now, in Eq.~\eqref{eq:skewdef}, we express $\tilde{\kappa}_{ij}$ as
\eq{
\hat{\tilde{\kappa}}_{ij} = \left( \hat{\tilde{\kappa}}^{-1} \right)^{-1}_{ij} \equiv \hat{U}^{-1}_{ij}
}
where, following Eq.~\eqref{eq:kappa2t},
\eq{
\hat{\tilde{\kappa}}^{-1}_{ij} = \frac{1}{VT^3} \, \left[\frac{1}{\alpha} \hat{\tilde{\chi}}^{'-1}_{ij} + \frac{1}{\beta} \hat{\tilde{\chi}}^{''-1}_{ij} \right].
}

The third order cumulant defined by Eq.~\eqref{eq:skewdef}, therefore, reads
\eq{\label{eq:kappa3st}
\hat{\tilde{\kappa}}_{i j k} = -\hat{\tilde{\kappa}}_{ij_1} \, \frac{1}{VT^3} \,\left[\frac{1}{\alpha} \frac{ \partial \hat{\tilde{\chi}}^{'-1}_{j_1j_2} }{\partial \hat{t}_k} + \frac{1}{\beta} \frac{ \partial \hat{\tilde{\chi}}^{''-1}_{j_1j_2} }{\partial \hat{t}_k} \right] \, \hat{\tilde{\kappa}}_{j_2j}~.
}

Let us now evaluate $\partial \hat{\tilde{\chi}}^{'-1}_{j_1j_2} / \partial \hat{t}_k$.
Using the identity \eqref{eq:invder} gives
\eq{
\frac{ \partial \hat{\tilde{\chi}}^{'-1}_{j_1j_2} }{\partial \hat{t}_k} 
= -\hat{\tilde{\chi}}^{'-1}_{j_1 m_1} \, \frac{ \partial \hat{\tilde{\chi}}_{m_1 m_2}^{'} }{\partial \hat{t}_k} \,  \hat{\tilde{\chi}}^{'-1}_{m_2 j_2}~,
}
which, after applying the chain rule to the middle term, turns into
\eq{
\frac{ \partial \hat{\tilde{\chi}}^{'-1}_{j_1j_2} }{\partial \hat{t}_k} 
= -\hat{\tilde{\chi}}^{'-1}_{j_1 m_1} \, 
\frac{ \partial \hat{\tilde{\chi}}_{m_1 m_2}^{'} }{\partial \hat{\bar{\mu}}^1_{m_3}} \, 
\frac{ \partial \hat{\bar{\mu}}^1_{m_3} }{\partial \den{}^1_{m_4}} \,  
\frac{\partial \den{}^1_{m_4}} {\partial \mean{\hat{Q}^1_{m_5}}} \, 
\frac{\partial \mean{\hat{Q}^1_{m_5}}} {\partial \hat{t}_k} \, 
\hat{\tilde{\chi}}^{'-1}_{m_2 j_2}~.
}
Using all the identities we derived above~[Eqs.~\eqref{eq:drho1dQ1},~\eqref{eq:dQ1dt}, and~\eqref{eq:dmudrho}] for each of the new terms, as well as the definition $\hat{\tilde{\chi}}_{ijk} = \partial \hat{\tilde{\chi}}_{ij} / \partial \hat{\bar{\mu}}_k$ of the third order GCE susceptibilities, we get
\eq{\label{eq:derinvchi1}
\frac{ \partial \hat{\tilde{\chi}}^{'-1}_{j_1j_2} }{\partial \hat{t}_k} 
= -\frac{1}{\alpha V T^3} \,
\hat{\tilde{\chi}}^{'-1}_{j_1 m_1} \, 
\hat{\tilde{\chi}}_{m_1 m_2 m_3}^{'} \, 
\hat{\tilde{\chi}}^{'-1}_{m_3 m_4} \,  
\hat{\tilde{\kappa}}_{m_4 k} \,
\hat{\tilde{\chi}}^{'-1}_{m_2 j_2}~.
}

The derivation for $\partial \hat{\tilde{\chi}}^{''-1}_{j_1j_2} / \partial \hat{t}_k$ that appears in the second term of the expression for $\hat{\tilde{\kappa}}_{i j k}$~[Eq.~\eqref{eq:kappa3st}] is analogous.
The only difference is the prefactor:
\eq{\label{eq:derinvchi2}
\frac{ \partial \hat{\tilde{\chi}}^{''-1}_{j_1j_2} }{\partial \hat{t}_k} = \frac{1}{\beta V T^3} \,
\hat{\tilde{\chi}}^{''-1}_{j_1 m_1} \, 
\hat{\tilde{\chi}}_{m_1 m_2 m_3}^{''} \, 
\hat{\tilde{\chi}}^{''-1}_{m_3 m_4} \,  
\hat{\tilde{\kappa}}_{m_4 k} \,
\hat{\tilde{\chi}}^{''-1}_{m_2 j_2}~.
}

The final result for $\hat{\tilde{\kappa}}_{i j k}(\hat{t})$ is obtained by substituting Eqs.~\eqref{eq:derinvchi1} and~\eqref{eq:derinvchi2} into~\eqref{eq:kappa3st}:
\eq{\label{eq:kappa3t}
\hat{\tilde{\kappa}}_{i j k}(\hat{t}) & =
\frac{1}{\alpha^2 V^2 T^6} \, 
\hat{\tilde{\kappa}}_{ij_1} \, 
\hat{\tilde{\chi}}^{'-1}_{j_1 m_1} \, 
\hat{\tilde{\chi}}^{'}_{m_1 m_2 m_3} \, 
\hat{\tilde{\chi}}^{'-1}_{m_3 m_4} \,  
\hat{\tilde{\kappa}}_{m_4 k} \,
\hat{\tilde{\chi}}^{'-1}_{m_2 j_2} \,
\hat{\tilde{\kappa}}_{j_2j} \nonumber \\
& \quad 
- \frac{1}{\beta^2 V^2 T^6}
\hat{\tilde{\kappa}}_{ij_1} \,
\hat{\tilde{\chi}}^{''-1}_{j_1 m_1} \, 
\hat{\tilde{\chi}}^{''}_{m_1 m_2 m_3} \, 
\hat{\tilde{\chi}}^{''-1}_{m_3 m_4} \,  
\hat{\tilde{\kappa}}_{m_4 k} \,
\hat{\tilde{\chi}}^{''-1}_{m_2 j_2} \,
\hat{\tilde{\kappa}}_{j_2j}~.
}

For $\hat{t} = 0$  we have $\hat{\tilde{\chi}}^{'} =
\hat{\tilde{\chi}}^{''} = \hat{\chi}$ for all ranks of the susceptibility tensor $\hat{\chi}$. Also  $\hat{\kappa}_{ij} = \alpha V T^3 \,  \beta \,
\hat{\chi}_{ij}$~[Eq.~\eqref{eq:kappa2final1}]. This implies that every convolution of $\hat{\kappa}_{ij_1}$ with $\hat{\chi}^{-1}_{j_1j}$ provides a Kronecker symbol times a factor, namely $\hat{\kappa}_{ij_1} \,\hat{\chi}^{-1}_{j_1j} = \delta_{ij} \, VT^3 \, \alpha \beta$.
This simplifies the evaluation of $\hat{\kappa}_{i j k}$ considerably.
In the end we obtain the following:
\eq{\label{eq:kappa3final1}
\hat{\kappa}_{ijk}[\hat{Q}^1] = \alpha V T^3 \,  \beta \, (1 - 2\alpha) \, \hat{\chi}_{ijk}~, \qquad i,j,k = 1,\ldots,N.
}

Similar to the second order cumulants~[Eq.~\eqref{eq:kappa2final1}], the  global conservation
corrections are identical for all third order cumulants. 
As a consequence, effects of global conservation cancel in any ratio of any two third order cumulants of conserved charges.

\subsection{Results  up to sixth order}

The fourth order and higher order cumulants are calculated by iteratively differentiating Eq.~\eqref{eq:kappa3t}. For example, for the fourth order cumulants one has:
\eq{\label{eq:kurtdef}
\hat{\tilde{\kappa}}_{i j k l}(\hat{t}) = \frac{\partial \hat{\tilde{\kappa}}_{ijk}(\hat{t})}{\partial \hat{t}_{l}}~, \qquad i,j,k,l = 1,\ldots,N.
}

The details of the iterative procedure for calculating higher-order cumulants are described in Appendix~\ref{app:highorder}.
Here we present the resulting SAM expressions up to the sixth order, evaluated at $\hat{t} = 0$.
For completeness we list here all cumulants starting from the first order:
\eq{
\label{eq:kappa1final}
\hat{\kappa}_{i_1}[\hat{Q}^1] & = \alpha V T^3 \, \hat{\chi}_{i_1}, 
\\
\label{eq:kappa2final}
\hat{\kappa}_{i_1 i_2}[\hat{Q}^1] & = \alpha V T^3 \, \beta \, \hat{\chi}_{i_1 i_2},
\\
\label{eq:kappa3final}
\hat{\kappa}_{i_1 i_2 i_3}[\hat{Q}^1] & = \alpha V T^3 \,  \beta \, (1 - 2\alpha) \, \hat{\chi}_{i_1 i_2 i_3} ,
\\
\label{eq:kappa4final}
\hat{\kappa}_{i_1 i_2 i_3 i_4}[\hat{Q}^1] & = \alpha V T^3 \,  \beta \left[
\, (1-3\alpha \beta) \, \hat{\chi}_{i_1i_2i_3i_4} - \frac{\alpha \beta}{2!\, 2!\, 2!} \sum_{\sigma \in S_4} \hat{\chi}^{-1}_{b_1b_2} \, \hat{\chi}_{\per{1}\per{2}b_1} \,  \hat{\chi}_{\per{3}\per{4}b_2} \, \right],
\\
\label{eq:kappa5final}
\hat{\kappa}_{i_1 \ldots i_5}[\hat{Q}^1] & = \alpha V T^3 \, \beta (1-2\alpha) 
\left[ 
(1-2\alpha \beta) \hat{\chi}_{i_1\ldots i_5}
- \frac{\alpha \beta}{2! \, 3!} \sum_{\sigma \in S_5} \hat{\chi}^{-1}_{b_1b_2} \, \hat{\chi}_{\per{1}\per{2}b_1} \,  \hat{\chi}_{\per{3}\per{4}\per{5}b_2}
\right],
}
\eq{
\label{eq:kappa6final}
\hat{\kappa}_{i_1 \ldots i_6}[\hat{Q}^1] & = \alpha  V T^3 \, \beta
\left\{
[1-5\alpha \beta(1-\alpha \beta)] \hat{\chi}_{i_1\ldots i_6} \right. \nonumber \\
& 
\quad + \frac{\alpha^2 \beta^2}{2!\,2!\,2!\,2!} \sum_{\sigma \in S_6} 
\hat{\chi}^{-1}_{b_1b_3} \, \hat{\chi}^{-1}_{b_2b_4} \,
\hat{\chi}_{\per{1}\per{2}b_1} \, \hat{\chi}_{\per{3}\per{4}b_2} \, 
\hat{\chi}_{\per{5}\per{6}b_3b_4} 
\nonumber \\
& \quad - \frac{\alpha^2 \beta^2}{3! \, 2! \, 2! \, 2!} 
\sum_{\sigma \in S_6} 
\hat{\chi}^{-1}_{b_1b_4} \, \hat{\chi}^{-1}_{b_2b_5} \, \hat{\chi}^{-1}_{b_3b_6} \,
\hat{\chi}_{b_4b_5b_6} \, \hat{\chi}_{\per{1}\per{2}b_1} \,  \hat{\chi}_{\per{3}\per{4}b_2} \, \hat{\chi}_{\per{5}\per{6}b_3} 
\nonumber \\
& \quad - \frac{\alpha \beta (1-2\alpha)^2}{2! \, 3! \, 3!} 
\sum_{\sigma \in S_6} 
\hat{\chi}^{-1}_{b_1b_2} \, \hat{\chi}_{\per{1}\per{2}\per{3}b_1} \, \hat{\chi}_{\per{4}\per{5}\per{6}b_2}
\nonumber \\
& \quad \left.
- \frac{\alpha \beta (1-3\alpha \beta)}{2! \, 4!} 
\sum_{\sigma \in S_6} 
\hat{\chi}^{-1}_{b_1b_2} \, \hat{\chi}_{\per{1}\per{2}b_1} \, \hat{\chi}_{\per{3}\per{4}\per{5}\per{6}b_2}
\right\}.
}
Here $i_1,\dots,i_6 = 1,\ldots,N$ in all of the above equations.
The notation $\sum_{\sigma \in S_M}$ corresponds to a sum over all M! permutations of a
set $(1,\ldots,M)$.
$\sigma_i$ is the $i$th element of the permutation $\sigma$.
In Appendix~\ref{app:illustr} we provide an example of a detailed calculation of a fourth order cumulant using Eq.~\eqref{eq:kappa4final}. This example is useful for understanding the notation entering Eqs.~\eqref{eq:kappa1final}-\eqref{eq:kappa6final}.
A \texttt{Mathematica} notebook to express cumulants in Eqs.~\eqref{eq:kappa1final}-\eqref{eq:kappa6final} in terms of the susceptibilities using the QCD notation~\eqref{eq:suscdefQCD} is available via~\cite{SAMgithub}.

As follows from Eqs.~\eqref{eq:kappa1final}-\eqref{eq:kappa3final},
the effects of global conservation laws and the equation of state factorize in cumulants up to third order:
these quantities are proportional to a product of the corresponding binomial~(Bernoulli) distribution cumulant and grand-canonical susceptibility.
For this reason, the global conservation factors cancel out in ratios of second-order cumulants and in ratios of third-order cumulants. 
As seen from Eq.~\eqref{eq:kappa4final}, this is no longer the case for fourth order cumulants: the global conservation effects generally affect different fourth order cumulants in a different way, thus
the global conservation effects do not cancel in ratios of fourth order cumulants.
An exception to this are vanishing chemical potentials, $\hat{\mu} = 0$, where $\hat{\chi}_{ijk} = 0$ for all $i,j,k$ and where only the first term in the r.h.s of~\eqref{eq:kappa4final} is non-zero.
Thus, at $\hat{\mu} = 0$ the effects of global conservation cancel in any ratio of any two four order cumulants.
The fifth order cumulants~\eqref{eq:kappa5final} have a structure similar to the fourth order cumulants.
The sixth order cumulants~\eqref{eq:kappa6final} have a considerably more involved structure.

Note that for a fixed value of $\alpha$, the cumulants $\hat{\kappa}_{i_1 \ldots i_M}[\hat{Q} - \hat{Q}^1]$ of conserved charge distribution in the second subsystem are obtained from Eqs.~\eqref{eq:kappa1final}-\eqref{eq:kappa3final} by a substitution $\alpha \to (1-\alpha)$~(or $\alpha \leftrightarrow \beta$).
One can observe that cumulants for the subsystem and the complement are related to each other via
\eq{\label{eq:complement}
\hat{\kappa}_{i_1 \ldots i_M}[\hat{Q}^1] = (-1)^M \, \hat{\kappa}_{i_1 \ldots i_M}[\hat{Q} - \hat{Q}^1], \qquad M \geq 2,
}
i.e. all even cumulants are equal between the two subsystems and all odd cumulants are opposite of each other.
The relation~\eqref{eq:complement} was rigorously derived in Ref.~\cite{Bzdak:2017ltv} for the case of single conserved charge. 
Here we observe it to hold true also for the case of multiple conserved charges, at least up to sixth order.
One consequence of Eq.~\eqref{eq:complement} is that all odd cumulants starting from third order vanish at $\alpha = 1/2$, as in this case the odd cumulants must not only be opposite to each other, as stipulated by Eq.~\eqref{eq:complement}, but also equal to one another due to symmetry.

\subsection{Conserved charges in QCD}
\label{sec:BQS}

\subsubsection{Single conserved charge $B$}

Let us consider the case of a single conserved charge -- the baryon number $B$. 
Then $\hat{Q} = (B)$, $N = 1$ and $\hat{\chi}_{i_1,\ldots,i_M} = \chi_M^B$, i.e. all $i_n = 1$.
Using the results of the preceding three subsections we can write the cumulants of baryon number $B^1$ inside a subvolume explicitly:
\eq{
\kappa_1[B^1] & = \alpha VT^3 \,  \chi_1^B, \\
\label{eq:kappa2B}
\kappa_2[B^1] & = \alpha VT^3 \, \beta \chi_2^B, \\
\kappa_3[B^1] & = \alpha VT^3 \, \beta (1-2\alpha) \chi_3^B, \\
\label{eq:kappa4Bsingle}
\kappa_4[B^1] & = \alpha VT^3 \, \beta \left[ 
 \, (1-3\alpha \beta) \, \chi_4^B
- 3 \, \alpha \, \beta \, \frac{(\chi_3^B)^2}{\chi_2^B}
\right], \\
\kappa_5[B^1] & = \alpha VT^3 \, \beta (1-2\alpha)  \left[ 
(1-2\alpha \beta)  \, \chi_5^B
- 10 \alpha \beta \, \frac{\chi_3^B \chi_4^B}{\chi_2^B}
\right], \\\nonumber
\kappa_6[B^1] & = \alpha VT^3 \, \beta \left\{ 
[1-5\alpha \beta(1-\alpha \beta)] \, \chi_6^B
+45 \alpha^2 \beta^2 \frac{(\chi_3^B)^2\chi_4^B}{(\chi_2^B)^2}
-15 \alpha^2 \beta^2 \frac{(\chi_3^B)^4}{(\chi_2^B)^3}
\right.\\
& \quad 
-10 \alpha \beta (1-2\alpha)^2 \, \frac{(\chi_4^B)^2}{\chi_2^B}-15 \left.\alpha \beta (1-3\alpha\beta) \, \frac{\chi_3^B \chi_5^B}{\chi_2^B}\right\}.
}
These expressions reproduce the results of Ref.~\cite{Vovchenko:2020tsr}, where the SAM was originally formulated for the case of a single conserved charge.

\subsubsection{Two conserved charges $B$ and $Q$}

In a case of two conserved charges, say baryon number $B$ and electric charge $Q$, we have $\hat{Q} = (B,Q)$ and $N = 2$.
Here we would like to illustrate how the cumulants of baryon number are affected by the presence of exact conservation of other conserved charges.
First, we note that, following Eqs.~\eqref{eq:kappa1final}, \eqref{eq:kappa2final}, and \eqref{eq:kappa3final}, the first three cumulants of baryon number $B^1$ have the same expression in the case of a single charge, i.e. they are unaffected by the presence of the conserved electric charge $Q$~(or any other additional exactly conserved quantity).

To evaluate the fourth order cumulant, $\kappa_4[B^1]$~[Eq.~\eqref{eq:kappa4final}], we need to compute the convolution in the second term of the r.h.s. of Eq.~\eqref{eq:kappa4final}, making use of the inverse matrix of second order susceptibilities.
Appendix~\ref{app:illustr} provides the details of this calculation.
The result is
\eq{\label{eq:kappa4Btwo}
\kappa_4[B^1] & = \alpha VT^3 \, \beta \, \left[ 
 \, (1-3\alpha \beta) \, \chi_4^B
- 3 \, \alpha \, \beta \, \frac{ (\chi_3^B)^2 \chi_2^Q - 2 \chi_{21}^{BQ} \chi_{11}^{BQ} \chi_3^B + (\chi_{21}^{BQ})^2 \chi_2^B }{\chi_2^B \chi_2^Q - (\chi_{11}^{BQ})^2}
\right].
}

It is evident from Eq.~\eqref{eq:kappa4Btwo} that the presence of a conserved electric charge influences the fourth order baryon number cumulant if there are baryon-electric charge correlations in the underlying equation of state.
This implies corrections to the relation~\eqref{eq:kappa4Bsingle} derived in Ref.~\cite{Vovchenko:2020tsr} for a system with a single conserved charge.
To elucidate these corrections we rewrite Eq.~\eqref{eq:kappa4Btwo} in the following form:
\eq{\label{eq:kappa4Btwob}
\kappa_4[B^1] & = \alpha VT^3 \, \beta \, \left[ 
 \, (1-3\alpha \beta) \, \chi_4^B
- 3 \, \alpha \, \beta \, \frac{ (\chi_3^B)^2}{\chi_2^B} \, \left( \frac{ 1 - 2 \frac{\chi_{21}^{BQ}
      \chi_{11}^{BQ}}{\chi_2^Q\chi_3^B} + \frac{(\chi_{21}^{BQ})^2
      \chi_2^B}{\chi_2^Q(\chi_3^B)^2}}{1 - \frac{(\chi_{11}^{BQ})^2}{\chi_2^B \, \chi_2^Q}} \right)
\right].
}
Effects of $BQ$-correlations can be sizable when the grand canonical off-diagonal susceptibilities are comparable to the diagonal ones.
On the other hand, corrections to Eq.~\eqref{eq:kappa4Bsingle} are small when $\chi_{11}^{BQ}, \chi_{21}^{BQ} \ll \chi_2^Q$ or when the matter-antimatter symmetry is small, i.e. for zero~(small) $\hat{\mu}$.
In the latter case, realized in heavy-ion collisions at LHC energies, all odd-order susceptibilities are (close to) zero, thus the second term in the r.h.s. of Eq.~\eqref{eq:kappa4Btwob} (nearly) vanishes.

We would also like to discuss a particular case where $BQ$-correlations are sizable but the effect of electric charge conservation on net baryon kurtosis is nevertheless small. 
This situation can be realized in the low-energy limit of heavy-ion collisions.
There, one approximates the QCD matter by a non-interacting gas of protons and neutrons, i.e. the production of antiparticles and mesons is neglected.
Neglecting the effects of Fermi statistics and nuclear clusters, the conserved charge susceptibilities here simply count the mean numbers of protons and neutrons, namely $\chi_2^B = \chi_3^B  \sim N_p + N_n$ and $\chi_2^Q = \chi_{11}^{BQ} = \chi_{21}^{BQ} \sim N_p$. 
Inserting these relations into the last term in Eq.~\eqref{eq:kappa4Btwob} we obtain for the low-energy limit
\eq{\label{eq:kappa4Blowenergy}
\kappa_4[B^1] & \approx \alpha VT^3 \, \beta \, \left[ 
 \, (1-3\alpha \beta) \, \chi_4^B
- 3 \, \alpha \, \beta \, \frac{ (\chi_3^B)^2}{\chi_2^B} \, \frac{ 1 - \frac{N_p}{N_p + N_n}}{1 - \frac{N_p}{N_p + N_n}}
\right], \nonumber \\
& = \alpha VT^3 \, \beta \, \left[ 
 \, (1-3\alpha \beta) \, \chi_4^B
- 3 \, \alpha \, \beta \, \frac{ (\chi_3^B)^2}{\chi_2^B} \right].
}
As one can see in Eq.~\eqref{eq:kappa4Blowenergy}, the net-baryon kurtosis in the low-energy limit reduces to expression~\eqref{eq:kappa4Bsingle} describing the case of a single conserved baryon number.
Thus, the influence of electric charge conservation on $\kappa_4[B^1]$ is expected to be negligible both in the high-energy and low-energy limits, provided that contributions of nuclear clusters to the partition function are negligible.
We explore the effect for intermediate energies in Sec.~\ref{sec:HRG}.

The fifth order cumulant, $\kappa_5[B^1]$, reads
\eq{
 \kappa_5[B^1] & = \alpha V T^3 \beta (1-2\alpha ) 
 \nonumber \\ & \quad \times
 \left[(1-2\alpha \beta ) \chi
   _5^B -10\alpha \beta \frac{  \chi
   _{31}^{{BQ}} ( \chi _2^B  \chi _{21}^{{BQ}}{ -
   }\chi _{11}^{{BQ}}{  }\chi _3^B{ )+ }\chi _4^B{ ( }\chi
   _2^Q{  }\chi _3^B{ - }\chi _{11}^{{BQ}}{  }\chi
   _{21}^{{BQ}}) }{\chi_2^B \chi_2^Q - (\chi_{11}^{BQ})^2}\right].
}
The discussion above on the behavior of $\kappa_4[B^1]$ largely applies to $\kappa_5[B^1]$ as well.

Explicit expression for $\kappa_6[B^1]$ can be obtained by expanding Eq.~\eqref{eq:kappa6final}. The expression is very lengthy, containing many different contributions from various $BQ$ correlators. 
Here we only present the explicit expression for $\kappa_6[B^1]$ at $\hat{\mu} = 0$, where all
odd-order susceptibilities vanish and thus simplify the formula considerably:
\eq{
 \left. \kappa_6[B^1] \right|_{\hat{\mu}=0} & = \alpha V T^3 \, \beta \left[(1-5\alpha \beta (1-\alpha \beta )) \chi_6^B \right . \nonumber \\
 & \quad \left. +10\alpha \beta (1-2\alpha )^2\frac{ \chi _2^B{ 
   (}\chi _{31}^{{BQ}})^2{ + }\chi _4^B{ ( }\chi _2^Q{ 
   }\chi _4^B{ -2 }\chi _{11}^{{BQ}}{  }\chi
   _{31}^{{BQ}} )}{\chi_2^B \chi_2^Q - (\chi_{11}^{BQ})^2}\right].
}

\subsubsection{Three conserved charges $B$, $Q$, and $S$}

Let us consider now three conserved charges: baryon number $B$, electric charge $Q$, and strangeness $S$.
In this case $\hat{Q} = (B,Q,S)$ and $N = 3$.
The first three cumulants of baryon number $B^1$ are the same as in the case of a single baryon charge.
The fourth cumulant reads
\eq{\label{eq:kappa4Bthree}
\kappa_4[B^1] & =  \alpha VT^3 \, \beta \left[ (1-3\alpha \beta) \, \chi_4^B - \frac{3 \alpha \beta}{D[\hat{\chi}_2]} \times \right. \nonumber \\
& \quad \left\{  (\chi_3^B)^2 [\chi_2^Q \chi_2^S - (\chi_{11}^{QS})^2] + (\chi_{21}^{BQ})^2 [\chi_2^B \chi_2^S - (\chi_{11}^{BS})^2] + (\chi_{21}^{BS})^2 [\chi_2^B \chi_2^Q - (\chi_{11}^{BQ})^2]  \right.
\nonumber \\
& \quad \left. \left. 
- 2 \chi_3^B \chi_{21}^{BQ} (\chi_2^S \chi_{11}^{BQ} - \chi_{11}^{BS} \chi_{11}^{QS})
- 2 \chi_3^B \chi_{21}^{BS} (\chi_2^Q \chi_{11}^{BS} - \chi_{11}^{BQ} \chi_{11}^{QS})
\right\} \right].
}
Here $D[\hat{\chi}_2]$ is the determinant of the matrix of second order 
susceptibilities:
\eq{\label{eq:Dchi2}
D[\hat{\chi}_2] = \chi_2^B \chi_2^Q \chi_2^S + 2 \chi_{11}^{BQ} \chi_{11}^{BS} \chi_{11}^{QS} - \chi_2^B \, (\chi_{11}^{QS})^2 - \chi_2^Q \, (\chi_{11}^{BS})^2 - \chi_2^S \, (\chi_{11}^{BQ})^2.
}

The fourth-order cumulant $\kappa_4[B^1]$ is affected by both the baryon-electric charge and baryon-strangeness correlations. 
Even the correlation between electric charge and strangeness does contribute, through a correlator $\chi_{11}^{QS}$.
It is notable that the entire second term in the r.h.s. of Eq.~\eqref{eq:kappa4Bthree} vanishes at
 LHC energies ($\hat{\mu}=0$), i.e.
\eq{
\kappa_4[B^1]|_{\hat{\mu}=0} = \alpha VT^3 \, \beta \, (1-3\alpha \beta) \, \chi_4^B~. 
}

We do not write here the lengthy expressions for $\kappa_5[B^1]$ and $\kappa_6[B^1]$.
These can be worked out from Eqs.~\eqref{eq:kappa5final} and \eqref{eq:kappa6final}, if desired.
We will only write, for completeness, the expression for $\kappa_6[B^1]$  for $\hat{\mu}=0$ (LHC energies), where it is considerably simplified:
\eq{\nonumber
\left. \kappa_6[B^1] \right|_{\hat{\mu}=0} & =\alpha VT^3 \, \beta \left[(1-5\alpha \beta (1-\alpha \beta
   ))\chi _6^B- \frac{10\alpha \beta (1-2\alpha
   )^2}{D[\hat{\chi}_2]}\times\right.\\\nonumber
   & \left\{(\chi _{31}^{{BS}})^2{[\chi_2^B \chi_2^Q - (\chi_{11}^{BQ})^2]+(}\chi
   _{31}^{{BQ}})^2[\chi_2^B \chi_2^S - (\chi_{11}^{BS})^2]\right.
   \\\nonumber
   & +(\chi _4^B)^2[\chi_2^Q \chi_2^S - (\chi_{11}^{QS})^2]
    +2\chi
   _{31}^{{BS}}\chi _{31}^{{BQ}}(\chi _{11}^{{BS}}\chi
   _{11}^{{BQ}}-\chi _2^B\chi _{11}^{{QS}}) \\
   & \left.\left.+
   2\chi_{31}^{{BS}}\chi _4^B(\chi _{11}^{{QS}}\chi
   _{11}^{{BQ}}-\chi _2^Q\chi _{11}^{{BS}}{)+2}\chi
   _{31}^{{BQ}}\chi _4^B(\chi _{11}^{{QS}}\chi
   _{11}^{{BS}}-\chi _2^B\chi _{11}^{{BQ}}\text)\right\}\right].
}

\subsection{Strongly intensive quantities}
\label{sec:SIQ}

Our considerations in the present paper are focused on effects of global conservation of multiple conserved charges, and how they distort the grand canonical baseline in the measured cumulants.
Another non-dynamical source that affects the measurements are fluctuations of the system volume, that cannot be completely avoided in heavy-ion collisions.
Different methods exist to 
address volume fluctuations~\cite{Gorenstein:2011vq,Skokov:2012ds,Braun-Munzinger:2016yjz}.
Here we explore briefly the possibility to construct quantities that are insensitive to both the global charge conservation and the volume fluctuations.

More specifically, we consider strongly intensive quantities -- fluctuation measures that were developed in Ref.~\cite{Gorenstein:2011vq} and designed to be insensitive to volume fluctuations.
These comprise of two combinations of first and second moments of two extensive quantities. 
Here we take two conserved charges, say $Q_a$ and $Q_b$, both measured in a subvolume $V_1$.
The strongly intensive quantities can be written in terms of cumulants as follows:
\eq{\label{eq:SIM}
\Delta[Q_a,Q_b] & = C_\Delta^{-1} \, \left\{ \kappa_1[Q_b] \frac{\kappa_2[Q_a]}{\kappa_1[Q_a]} - \kappa_1[Q_a] \frac{\kappa_2[Q_b]}{\kappa_1[Q_b]} \right\}, \\
\Sigma[Q_a,Q_b] & = C_\Sigma^{-1} \, \left\{ \kappa_1[Q_b] \frac{\kappa_2[Q_a]}{\kappa_1[Q_a]} + \kappa_1[Q_a] \frac{\kappa_2[Q_b]}{\kappa_1[Q_b]} - 2 \, \kappa_{1,1}[Q_a,Q_b] \right\}.
}

The normalization factors $C_\Delta$ and $C_\Sigma$ correspond to an arbitrary extensive measure not sensitive to volume fluctuations. 
As an example, one can take any linear combination of mean values of $Q_a$ and $Q_b$.
Possible specific choices of $C_\Delta$ and $C_\Sigma$ have been discussed in Ref.~\cite{Gazdzicki:2013ana}.

The quantities $\Delta[Q_a,Q_b]$ and $\Sigma[Q_a,Q_b]$~(and any combination of the two) are insensitive to fluctuations of the total system volume $V$.
To show their sensitivity to global charge conservation we use Eqs.~\eqref{eq:kappa1final} and \eqref{eq:kappa2final}:
\eq{
\Delta[Q_a,Q_b] & = C_\Delta^{-1} \, VT^3 \, \alpha (1-\alpha) \left\{ \chi_1^{Q_b} \frac{\chi_2^{Q_a}}{\chi_1^{Q_a}} - \chi_1^{Q_a} \frac{\chi_2^{Q_b}}{\chi_1^{Q_b}} \right\}, \\
\Sigma[Q_a,Q_b] & = C_\Sigma^{-1} \, VT^3 \, \alpha (1-\alpha) \left\{ \chi_1^{Q_b} \frac{\chi_2^{Q_a}}{\chi_1^{Q_a}} + \chi_1^{Q_a} \frac{\chi_2^{Q_b}}{\chi_1^{Q_b}} - 2 \, \chi_{1,1}^{Q_a Q_b}\right\}.
}
We note that the above two equations are obtained assuming that the subvolume fraction $\alpha$ is constant, i.e. it is unaffected by volume fluctuations.
If $\alpha$ does fluctuate, e.g. due to possible fluctuations in baryon stopping, the formalism will require a generalization to account for that.

Given that $C_\Delta$ and $C_\Sigma$ are proportional to any extensive~(i.e. proportional to the subvolume $V_1$) measure that is not sensitive to volume fluctuations, we can, without the loss of generality, write these factors as $C_\Delta = \alpha V T^3 \chi_1^\Delta$ and $C_\Sigma = \alpha V T^3 \chi_1^\Sigma$ where $\chi_1^\Delta$ and $\chi_1^\Sigma$ do not depend on $V$ and $\alpha$, e.g. they may be chosen as linear combinations of $\chi_1^{Q_a}$ and $\chi_1^{Q_b}$.
Then:
\eq{
\Delta[Q_a,Q_b] & = \frac{1-\alpha}{\chi_1^\Delta} \left\{ \chi_1^{Q_b} \frac{\chi_2^{Q_a}}{\chi_1^{Q_a}} - \chi_1^{Q_a} \frac{\chi_2^{Q_b}}{\chi_1^{Q_b}} \right\}, \\
\Sigma[Q_a,Q_b] & = \frac{1-\alpha}{\chi_1^\Sigma} \left\{ \chi_1^{Q_b} \frac{\chi_2^{Q_a}}{\chi_1^{Q_a}} + \chi_1^{Q_a} \frac{\chi_2^{Q_b}}{\chi_1^{Q_b}} - 2 \, \chi_{1,1}^{Q_a Q_b}\right\}.
}

Both strongly intensive measures $\Delta[Q_a,Q_b]$ and $\Sigma[Q_a,Q_b]$ are affected by global
charge conservation, through a common factor $1-\alpha$. This implies that a ratio of these two
quantities is neither affected by  volume fluctuations nor the global charge conservation:
\eq{
\frac{\Sigma[Q_a,Q_b]}{\Delta[Q_a,Q_b]} = \frac{\chi_1^\Delta}{\chi_1^\Sigma} \, \frac{\chi_1^{Q_b} \ddfrac{\chi_2^{Q_a}}{\chi_1^{Q_a}} + \chi_1^{Q_a} \ddfrac{\chi_2^{Q_b}}{\chi_1^{Q_b}} - 2 \, \chi_{1,1}^{Q_a Q_b}}{\chi_1^{Q_b} \ddfrac{\chi_2^{Q_a}}{\chi_1^{Q_a}} - \chi_1^{Q_a} \ddfrac{\chi_2^{Q_b}}{\chi_1^{Q_b}}}~.
}

\subsection{Non-conserved quantities}
\label{sec:SEnoncons}

So far we focused on the cumulants of multiple conserved charges. 
It is, however, notoriously difficult to measure neutral particles event-by-event in heavy-ion experiments, preventing direct measurements of baryon number and strangeness fluctuations.
For this reason one usually considers non-conserved quantities such as net-proton or net-kaon number as proxies for net-baryon and net-strangeness fluctuations.
The measurements of electric charge fluctuations do not suffer from this problem and can be 
done directly.
For instance, the STAR collaboration has recently reported measurements of net-proton, net-kaon, and net-charge second order cumulants~\cite{Adam:2019xmk}.

Here we consider the behavior of a non-conserved quantity, such as net-proton or net-kaon number, in the presence of exact conservation of 
charges.
For clarity, we will refer to this quantity as a net proton number, $N_p$. 
The considerations below, however, are general and apply to any non-conserved quantity, not only net-proton number.

As the net-proton number is not a conserved quantity, it has a distribution for fixed values of the conserved charges $\hat{Q}$. This implies that the canonical partition function can be written as sum over all possible values of $N_p$
\eq{\label{eq:ZNp}
Z(T,V,\hat{Q}) = \sum_{N_p} \, W(T,V,\hat{Q};N_p)~.
}
Here $W(T,V,\hat{Q};N_p)$ counts the number of configurations that yield a particular net-proton
number $N_p$ in the final state. 
We note that $W(T,V,\hat{Q};N_p)$ is not easily accessible in theoretical calculations, 
such as lattice QCD, as it requires a careful projection on asymptotic states that count the net number of protons emerging from a thermalized QCD matter created in heavy-ion collision. 
However, the number of (anti)protons in a given heavy-ion event is a well
defined, physical observable accessible to experiment. 
In the framework of the hadron resonance gas (HRG), 
which is successfully applied to extract the chemical freeze-out
temperature, 
the number of protons is given by the sum of primordial protons and those arising from all strong and electromagnetic
decays of resonances, such as for example $\Delta(1232)$. 
Or in other words, the projection to asymptotic states
in the HRG is equivalent to taking all resonance decays into account. 
For the purposes of this paper is not important how and if $W(T,V,\hat{Q};N_p)$ can be calculated in a given theory but rather that it is a well defined quantity, which it is.
Since $N_p$ is not conserved, the system can freely fluctuate from a configuration with a particular
value of $N_p$ to a configuration with another value.
Thus, $N_{p}$ behaves just as a conserved charge in a grand canonical ensemble, so that $W(T,V,\hat{Q};N_p)$ can be regarded as a generalized canonical partition function with fixed values of conserved charges $\hat{Q}$ and net-proton number $N_p$.
In the thermodynamic limit, $V \to \infty$, the following representation of $W(T,V,\hat{Q};N_p)$ holds:
\eq{\label{eq:WNp}
W (T,V,\hat{Q};N_p) = \exp\left[-\frac{V}{T} \, \breve{f}(T,\dens{},\rho_p) \right].
}
Here $\breve{f}(T,\dens{},\rho_p)$ is a generalized free energy density that depends on net-proton density $\rho_p = N_p / V$, in addition to the conserved charge densities.
In the thermodynamic limit $W(T,V,\hat{Q};N_p)$ is highly peaked around $\mean{N_p}$, the sum~\eqref{eq:ZNp} is determined by the maximum term at $N_p = \mean{N_p}$, i.e.
\eq{
f(T,\dens{}) & \stackrel{V \to \infty}{=} \breve{f}(T,\dens{},\rho_p^0)~,
}
where $\rho_p^0 = \mean{N_p} / V$.

By considering the canonical partition function $Z(T,V,\hat{Q})$ in the form given by Eqs.~\eqref{eq:ZNp} and \eqref{eq:WNp} we can now introduce a generalized cumulant generating function for the joint distribution of $\hat{Q}^1$, $N_p^1$, and $N_p^2$.
Here $N_p^1$ and $N_p^2$ are the net-proton numbers in the first and second subsystems, respectively.
The generating function reads
\eq{
\tilde{G}_{\hat{Q}^1,N_p^1,N_p^2}(\hat{t},t_p^1,t_p^2) & = \ln \left\{ \sum_{\hat{Q}^1,N_p^1,N_p^2} \, e^{\hat{t}_i \hat{Q}^1_i + t_p^1 N_p^1 + t_p^2 N_p^2} \, 
e^{-\frac{\alpha V}{T} \, \breve{f}(T,\dens{}^1,\rho_p^1)}
e^{-\frac{\beta V}{T} \, \breve{f}(T,\dens{}^2,\rho_p^2)}
\right\} 
+ \tilde{C}~.
}
Derivatives of $\tilde{G}_{\hat{Q}^1,N_p^1,N_p^2}$ evaluated at $\hat{t} = 0$, $t_p^1 = t_p^2 = 0$ give mixed cumulants of the $(\hat{Q}^1,N_p^1,N_p^2)$ distribution.
One can see that $\tilde{G}_{\hat{Q}^1,N_p^1,N_p^2}(\hat{t},0,0) = G_{\hat{Q}^1}(\hat{t})$, therefore all cumulants that involve the conserved charges $\hat{Q}$, but not $N_p^{1,2}$, will coincide with the results obtained in previous section.

To evaluate the cumulants involving $N_p^1$ we maximize the generalized probability function
\eq{
\tilde{P}(\hat{Q}^1,N_p^1,N_p^2;\hat{t},t_1,t_2) = \exp\left\{\hat{t}_i \hat{Q}^1_i + t_p^1 N_p^1 + t_p^2 N_p^2 - V \, \frac{\alpha \breve{f}(T,\dens{}^1,\rho_p^1) + \beta \breve{f}(T,\dens{}^2,\rho_p^2)}{T}\right\}~
}
with respect to $N_p^1$:
\eq{\label{eq:tp1}
t_p^1 = \breve{\bar{\mu}}_p [T,\dens{}^1,\rho_p^1]~.
}
Here $\breve{\bar{\mu}}_p [T,\dens{}^1,\rho_p^1] = T^{-1} \, \partial \breve{f} / \partial \rho_p^1$.

\subsubsection{Off-diagonal cumulants involving a single conserved charge}

Let us introduce a matrix of the grand canonical second order susceptibilities for the joint $(\hat{Q},N_p)$ distribution:
\eq{\label{eq:tildechi}
\breve{\chi} = 
\begin{pmatrix}
 \chi_{\hat{Q}_i\hat{Q}_j} & \chi_{\hat{Q}_ip} \\
 \chi_{p\hat{Q}_j} & \chi_{pp}
\end{pmatrix},
\qquad i,j = 1,\ldots,N.
}
The $\breve{\chi}_{ij}$ matrix is $(N+1)\times(N+1)$ dimensional, and 
$\chi_{\hat{Q}_i\hat{Q}_j}$ is the $N \times N$ matrix of second order susceptibilities of conserved charges $\hat{Q}$, defined in Eq.~\eqref{eq:chi2def}, $\chi_{\hat{Q}_i p} = \chi_{p \hat{Q}_i} = \partial \rho_p / \partial \hat{\bar{\mu}}_i$ is a  grand canonical correlator between $N_p$ and conserved charge $\hat{Q}_i$, and $\chi_{pp} = \partial \rho_p / \partial \breve{\bar{\mu}}_p$ is the grand canonical susceptibility for the net-proton number $N_p$.

Differentiating Eq.~\eqref{eq:tp1} with respect to $\hat{t}_j$, where $j \in 1 \ldots N$,
we get
\eq{\label{eq:EqforpQ}
0 =  
\sum_{j_1=1}^N \left[\breve{\tilde{\chi}}^{-1}_{N+1,j_1} \tilde{\kappa}_{j_1 j}\right] + \breve{\tilde{\chi}}^{-1}_{N+1,N+1} \tilde{\kappa}_{N+1, j}~.
}
The tilde in $\breve{\tilde{\chi}}$ means that the susceptibilities are calculated in the first subsystem at arbitrary values of $\hat{t}$ and $t_p^1$, whereas the same quantity without a tilde corresponds to the susceptibilities evaluated at $\hat{t} = 0$ and $t_p^1 = 0$.
The same notation applies to the use of tilde in the notation for the cumulants $\kappa$ of the $(\hat{Q}^1,N_p^1)$ distribution.
$\breve{\tilde{\chi}}^{-1}_{N+1,j_1}$ and $\breve{\tilde{\chi}}^{-1}_{N+1,N+1}$ in Eq.~\eqref{eq:EqforpQ} correspond to the elements of the inverse $\breve{\tilde{\chi}}$ matrix~[Eq.~\eqref{eq:tildechi}].
$\tilde{\kappa}_{N+1, j}$ corresponds to a mixed cumulant involving net-proton number $N_p^1$ and conserved charge $\hat{Q}_j^1$, both evaluated in the first subsystem.

Equation \eqref{eq:EqforpQ} can be solved for $\tilde{\kappa}_{N+1, j}$, yielding:
\eq{
\tilde{\kappa}_{N+1, j} = -\frac{\sum_{j_1=1}^N \breve{\tilde{\chi}}^{-1}_{N+1,j_1} \tilde{\kappa}_{j_1 j}}{\breve{\tilde{\chi}}^{-1}_{N+1,N+1}}~.
}
Now we set $\hat{t} = 0$ and $t_p^1 = 0$, so that  $\breve{\tilde{\chi}}^{-1} \rightarrow \breve{\chi}^{-1}$.
Furthermore, for $\kappa_{j_1 j}$ with $j_1,j = 1\ldots N$ we can use the
result~\eqref{eq:kappa2final} leading to
\eq{\label{eq:pQ1}
\kappa_{N+1, j} = -\alpha  VT^3 \, \beta \, \frac{\sum_{j_1=1}^N \breve{\chi}^{-1}_{N+1,j_1} \breve{\chi}_{j_1 j}}{\breve{\chi}^{-1}_{N+1,N+1}}~.
}

We observe that $\sum_{j_1=1}^{N+1} \breve{\chi}^{-1}_{N+1,j_1} \breve{\chi}_{j_1 j} = 0$ as this expression corresponds to an off-diagonal element of the $(N+1)\times(N+1)$ identity matrix $\hat{I} = \breve{\chi}^{-1} \breve{\chi}$.
This implies $\sum_{j_1=1}^N \breve{\chi}^{-1}_{N+1,j_1} \breve{\chi}_{j_1 j} =
-\breve{\chi}^{-1}_{N+1,N+1} \breve{\chi}_{N+1, j}$ in Eq.~\eqref{eq:pQ1} so that
\eq{\label{eq:pQ2}
\kappa_{N+1, j} = \alpha VT^3 \, \beta  \, \breve{\chi}_{N+1, j}~.
}

Given the structure of $\breve{\chi}$~[Eq.~\eqref{eq:tildechi}] we can identify $\breve{\chi}_{N+1, j} = \chi_{p\hat{Q}_j}$ as a grand canonical correlator of net-proton number with a conserved charge $\hat{Q}_j$.
The corresponding off-diagonal cumulant $\kappa_{p\hat{Q}_j}$ reads
\eq{\label{eq:pQfinal}
\kappa_{p\hat{Q}_j} = \alpha VT^3 \, \beta  \, \chi_{p\hat{Q}_j}~.
}

The cumulant $\kappa_{p\hat{Q}_j}$ is affected by the global conservation of 
charges in the same way as all second order cumulants of conserved charges~[Eq.~\eqref{eq:kappa2final}].
This means effects of conservation laws cancel out in
following ratios:
\eq{\label{eq:Npratios}
\frac{\kappa_{p\hat{Q}_j}}{\kappa_{\hat{Q}_i\hat{Q}_j}} = \frac{\chi_{p\hat{Q}_j}}{\chi_{\hat{Q}_i\hat{Q}_j}}~,
\qquad
\frac{\kappa_{p\hat{Q}_j}}{\kappa_{p\hat{Q}_i}} = \frac{\chi_{p\hat{Q}_j}}{\chi_{p\hat{Q}_i}}~.
}
Furthermore, as our derivation has been obtained for an arbitrary non-conserved quantity,
we can also consider correlators of the electric charge with different non-conserved quantities, such as e.g. net proton and net kaon numbers.
The global conservation factors cancel
\eq{
\frac{\kappa_{p\hat{Q}_j}}{\kappa_{k\hat{Q}_i}} = \frac{\chi_{p\hat{Q}_j}}{\chi_{k\hat{Q}_i}}~.
}

\subsubsection{Variance of a non-conserved quantity}

Let us now differentiate Eq.~\eqref{eq:tp1} with respect to $t_p^1$:
\eq{
1 = \frac{1}{\alpha VT^3} 
\left\{  
\sum_{j_1=1}^N \left[\breve{\tilde{\chi}}^{-1}_{N+1,j_1} \tilde{\kappa}_{j_1,N+1}\right] + \breve{\tilde{\chi}}^{-1}_{N+1,N+1} \tilde{\kappa}_{N+1, N+1}
\right\}.
}
This can be solved for $\tilde{\kappa}_{N+1, N+1}$.
For $\hat{t} = 0$, $t_p^1 = 0$, the solution reads
\eq{
\kappa_{N+1, N+1} = \frac{\alpha VT^3}{\breve{\chi}^{-1}_{N+1,N+1}} 
\left[ 
1 - \beta\sum_{j_1=1}^N \breve{\chi}^{-1}_{N+1,j_1} \breve{\chi}_{j_1,N+1}
\right],
}
where we used the result~\eqref{eq:pQ2} for $\kappa_{j_1, N+1}$.
The sum $\sum_{j_1=1}^{N+1} \breve{\chi}^{-1}_{N+1,j_1} \breve{\chi}_{j_1,N+1}$ corresponds to 
the $(N+1),(N+1)$  element of an identity matrix $\hat{I} = \breve{\chi}^{-1} \breve{\chi}$, therefore, $\sum_{j_1=1}^N \breve{\chi}^{-1}_{N+1,j_1} \breve{\chi}_{j_1 N+1} = 1-\breve{\chi}^{-1}_{N+1,N+1} \breve{\chi}_{N+1, N+1}$~.
We, thus, obtain
\eq{
\kappa_{N+1, N+1} = 
\alpha \beta VT^3 \breve{\chi}_{N+1, N+1} + 
\frac{\alpha^2 VT^3}{\breve{\chi}^{-1}_{N+1,N+1}}~.
}
Given that $\breve{\chi}^{-1}_{N+1,N+1}$ is the lower right element of the inverse of matrix $\breve{\chi}$~[Eq.~\eqref{eq:tildechi}], it can be expressed as
$\breve{\chi}^{-1}_{N+1,N+1} = \frac{\det{\breve{\chi}}}{\det{\chi}}$
where $\chi=\chi_{\hat{Q}_i\hat{Q}_j}$ is a $N\times N$ matrix of grand canonical susceptibilities of conserved charges $\hat{Q}$.
The net-proton susceptibility reads
\eq{\label{eq:kappapp}
\kappa_{pp} & = 
\alpha VT^3 \left[ (1-\alpha) \chi_{pp} +\alpha \frac{\det{\breve{\chi}}}{\det{\chi}} \right].
}
By definition, in the limit $\alpha \to 1$ $\kappa_{pp}$ reduces to the variance of net-proton
distribution in the canonical ensemble $\kappa_{pp}^{\rm ce} \equiv VT^3 \chi_{pp}^{\rm
  ce}$. Therefore, by setting $\alpha = 1$ in Eq.~\eqref{eq:kappapp} we obtain the canonical net-proton susceptibility:
\eq{\label{eq:chippce}
\chi_{pp}^{\rm ce} =  \frac{\det{\breve{\chi}}}{\det{\chi}}~.
}

From an analysis of Eqs.~\eqref{eq:kappapp} and~\eqref{eq:chippce} one observes that the net-proton susceptibility $\kappa_{pp} / (\alpha VT^3)$  represents a linear combination of the grand canonical net-proton susceptibility $\chi_{pp}$ and the canonical net-proton susceptibility $\chi_{pp}^{\rm ce} = \frac{\det{\breve{\chi}}}{\det{\chi}}$ at all values of $\alpha$.
Note that $\chi_{pp}^{\rm ce}$ vanishes if the net-proton number coincides with one of the conserves charges, as in this case the matrix $\tilde{\chi}$ will have at least two identical rows~[see Eq.~\eqref{eq:tildechi}], thus its determinant is zero.
In this case Eq.~\eqref{eq:kappapp} reduces to the second cumulant of a conserved charge~[Eq.~\eqref{eq:kappa2final}], as it should  by construction.

\section{Application to the hadron resonance gas model}
\label{sec:HRG}

\subsection{HRG model setup}

To illustrate the main features of the SAM 
in the presence of multiple conserved charges we shall consider the behavior of various fluctuation measures in a hadron resonance gas~(HRG) model~\cite{Hagedorn:1965st}.
The HRG model describes the hadronic phase as a multi-component gas of known hadrons and resonances and has broad applications for describing hadrochemistry in relativistic heavy-ion collisions~\cite{Letessier:2005qe,Becattini:2009sc,Andronic:2017pug}.

In the present work we take the simplest variant of the HRG model where we neglect quantum statistics, finite resonance widths, and excluded volume corrections.
Even though the model in this case reduces simply to a multi-component ideal gas of Maxwell-Boltzmann particles, the different hadron species do carry different values of the three QCD conserved charges, baryon number $B$, electric charge $Q$, and strangeness $S$.
This induces non-trivial cross-correlations between conserved charges, making the model suitable for studying the effects of multiple global conservation laws.

For an arbitrary grand canonical HRG model the conserved charge susceptibilities are
\eq{\label{eq:chiHRG}
\chi^{BQS}_{lmn} & \equiv \frac{\partial^{l+m+n} \, p/T^4}{\partial \left(\mu_B / T\right)^l \partial \left(\mu_Q / T\right)^m \partial \left(\mu_S / T\right)^n} \nonumber \\
& = \sum_i \, (b_i)^l \, (q_i)^m \, (s_i)^n \, \frac{d_i}{2\pi^2} \, \left( \frac{m_i}{T} \right)^2 \, K_2\left( m_i \over T \right) \, \exp\left(
\frac{b_i \mu_B + q_i \mu_Q + s_i \mu_S}{T}\right)~ \nonumber \\
& = \sum_i \, (b_i)^l \, (q_i)^m \, (s_i)^n \, \chi_i^{\rm hrg}~, \\
\label{eq:chiiHRG}
\chi_i^{\rm hrg} & = \frac{d_i}{2\pi^2} \, \left( \frac{m_i}{T} \right)^2 \, K_2\left( m_i \over T \right) \, \exp\left(
\frac{b_i \mu_B + q_i \mu_Q + s_i \mu_S}{T}\right).
}
Here we used the conventional notation~[Eq.~\eqref{eq:suscdefQCD}] for the susceptibilities.
The index $i$ sums over all hadron species in the HRG, including both particles and
antiparticles. The degeneracy and mass of a hadron specie $i$ are denoted by $d_i$ and $m_i$, respectively, while $b_i$, $q_i$, and $s_i$ are its baryon number, electric charge, and strangeness.
We also introduced a shorthand $\chi_i^{\rm hrg}$~\eqref{eq:chiiHRG} for the susceptibility of particle number of hadron species $i$.
We take into account all established light-flavor and strange hadrons and resonances from the 2014 edition of Particle Data Tables~\cite{Agashe:2014kda}, as incorporated in the default particle list of the open source thermal-statistical package \texttt{FIST}~\cite{Vovchenko:2019pjl}.

For our calculation we use the following values of the thermal parameters: $T = 160$~MeV and $\mu_B = 100$~MeV.
The temperature value is typical for chemical freeze-out in heavy-ion
collisions~\cite{Andronic:2017pug,Adamczyk:2017iwn}, while a non-zero value of $\mu_B$ allows to
study effects that are absent at $\mu_B = 0$, 
such as mixing of cumulants of various order discussed in
Sec.~\ref{sec:BQS}.
With this choice, the HRG model approximates the chemical freeze-out conditions encountered in heavy-ion collisions at $\sqrt{s_{_{NN}}} \simeq 40-45$~GeV~\cite{Cleymans:2005xv,Vovchenko:2015idt}.
The electric charge and strangeness chemical potentials are fixed to yield an electric-to-baryon charge ratio $Q/B = 0.4$ and a vanishing net-strangeness $S = 0$, resulting in $\mu_Q \simeq -3$~MeV and $\mu_S \simeq 23$~MeV.
All the grand canonical conserved charge susceptibilities are then calculated using Eq.~\eqref{eq:chiHRG} while the results of Sec.~\ref{sec:subensemble} are used to establish the acceptance ($\alpha$) dependence of the cumulants in the presence of global conservation of baryon number, electric charge, and strangeness.

In addition to analytic calculations of cumulants within the SAM, we also perform Monte Carlo calculations of the same cumulants by sampling the HRG in the canonical ensemble.
We employ the canonical ensemble sampler of the \texttt{FIST} package to generate hadron multiplicities in the full acceptance.
The canonical ensemble sampler follows the efficient multi-step procedure introduced by Becattini
and Ferroni~\cite{Becattini:2004rq}, and its implementation in \texttt{FIST} is detailed in Ref.~\cite{Vovchenko:2018cnf}.
Given that hadron coordinates are uncorrelated in the HRG model, we then independently apply a Bernoulli trial with probability $\alpha$ to each hadron in each event to establish whether it belongs to a subvolume $V_1 = \alpha V$.
In our sampling procedure we take values $B = 20$, $Q = 8$, $S = 0$ of the globally conserved charges.
The value of the total volume of $V$ is calculated using the grand canonical HRG model such that the mean baryon number, electric charge, and strangness are equal to the canonical ensemble values listed above.
The resulting value of $V = 522.8$~fm$^3$ ensures consistency between the grand canonical and canonical formulations regarding all average quantities.
This value of $V$ is, on one hand, large enough such that deviations from the thermodynamic limit are small, and on the other hand, it is small enough to obtain sufficient statistics for an accurate calculation of cumulants up to the fourth order.
In total we generate $10^8$ events for the present analysis.

The Monte Carlo calculations present a powerful validation of the SAM, as they make no use of the assumptions which go into that formalism.

\subsection{Second order cumulants of conserved charges}

\begin{figure}[t]
  \centering
  \includegraphics[width=.49\textwidth]{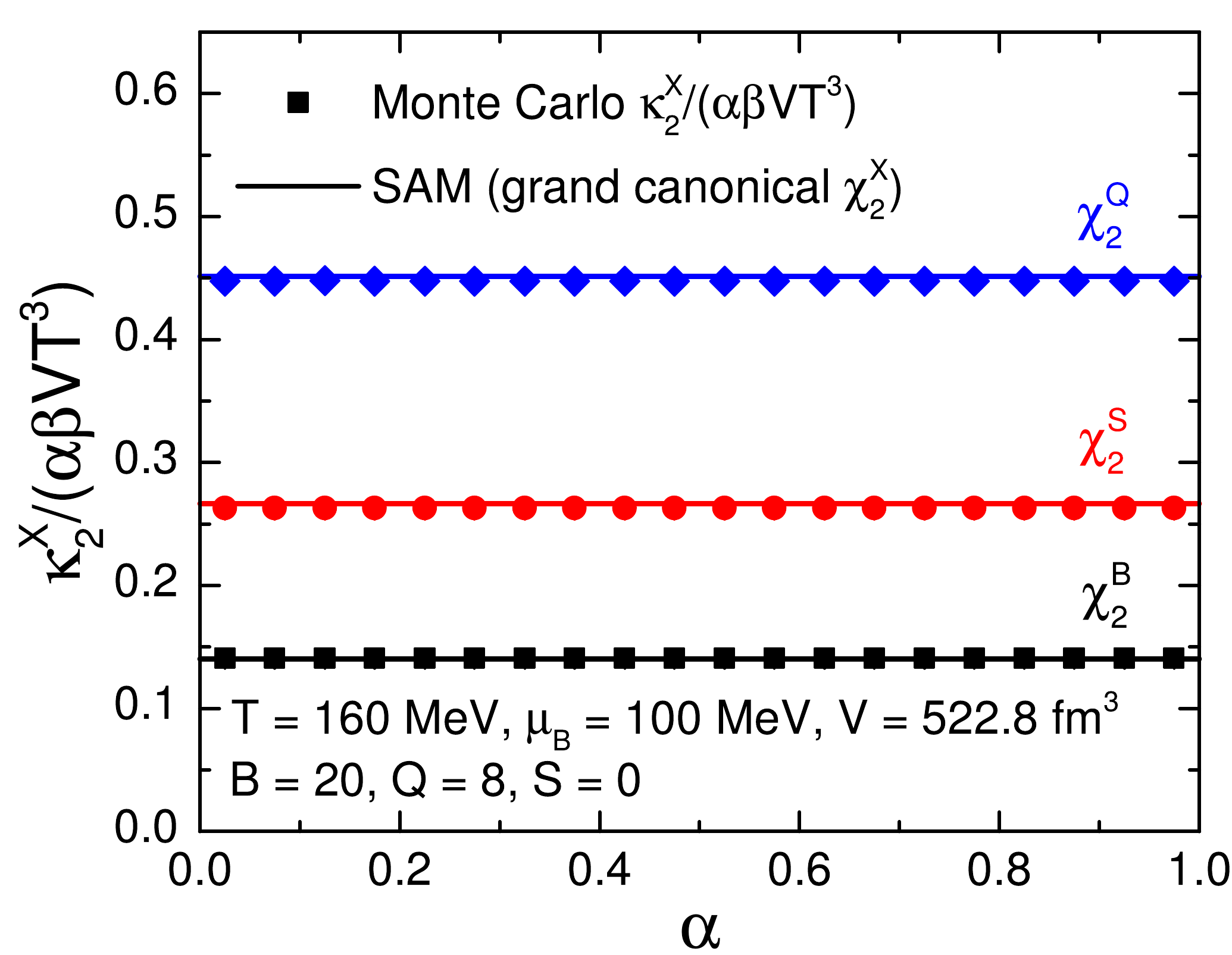}
  \includegraphics[width=.49\textwidth]{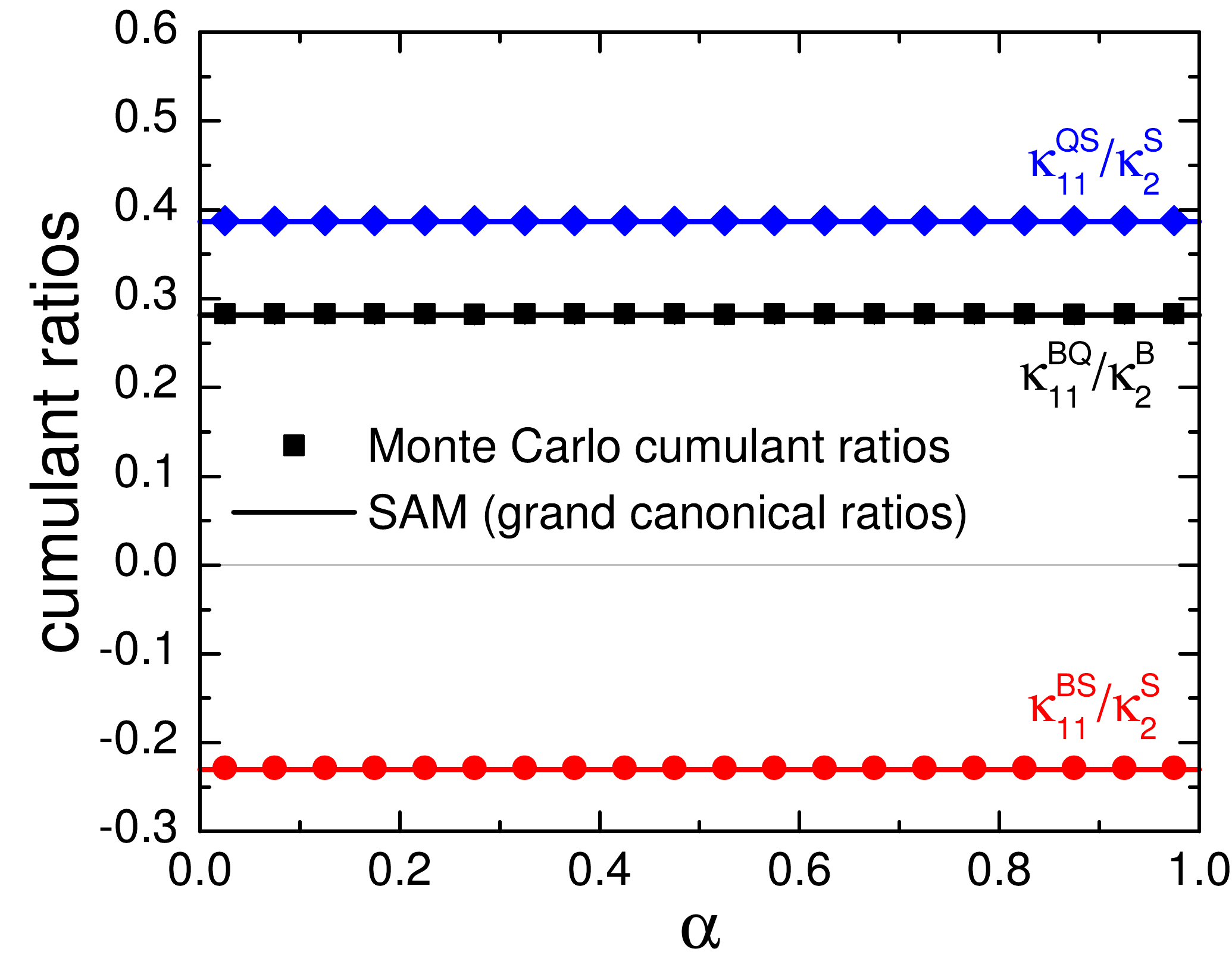}
  \caption{
  Dependence of combinations of second order cumulants of conserved charges on the acceptance fraction $\alpha$, as calculated in the hadron resonance gas model using canonical ensemble Monte Carlo sampler~(symbols) and analytically in the framework of the subensemble acceptance method~(lines).
  The Monte Carlo calculation contains $10^8$ events.
  \emph{Left panel:} Diagonal cumulants of net-baryon~(black), net-charge~(blue), and net-strangeness~(red) numbers scaled by a factor $\alpha \beta V T^3$, yielding the grand-canonical susceptibilities in the SAM~[Eq.~\eqref{eq:kappa2final}]. Here $\beta\equiv1-\alpha$.
  \emph{Right panel:} Off-diagonal to diagonal conserved charge cumulant ratios $\kappa_{11}^{BQ} / \kappa_2^B$~(black), $\kappa_{11}^{QS} / \kappa_2^S$~(blue), and $\kappa_{11}^{BS} / \kappa_2^S$~(red).
  }
  \label{fig:HRGchi2}
\end{figure}

The SAM predicts that appropriately scaled second order cumulants $\kappa_2^X / (\alpha \beta VT^3)$, where $X = B,Q,S$, are independent of the value of $\alpha$ and coincide with the corresponding grand canonical susceptibilities $\chi_2^X$~[see Eq.~\eqref{eq:kappa2final}].
The Monte Carlo calculations do indeed confirm this behavior, as shown in the left panel of Fig.~\ref{fig:HRGchi2}.
Measurements of the variance of conserved charges can thus be used to extract the grand canonical susceptibility if the values of the acceptance fraction $\alpha$, the volume $V$, and the temperature $T$ can be reliably estimated.

In practice, a reliable estimation of $\alpha$, $V$, and $T$ is challenging.
For this reason it is useful to consider observables where these quantities do not appear.
As already discussed, Eq.~\eqref{eq:kappa2final}, a ratio of any two second order cumulants of conserved charges is insensitive to $\alpha$ and $V$, and coincides with the corresponding ratio of grand canonical susceptibilities.
The right panel of Fig.~\ref{fig:HRGchi2} depicts the $\alpha$-dependence of cumulants ratios $\kappa_{11}^{BQ} / \kappa_2^B$, $\kappa_{11}^{QS} / \kappa_2^S$, and $\kappa_{11}^{BS} / \kappa_2^S$, as calculated within the Monte Carlo event generator~(symbols) and within the SAM using the grand canonical susceptiblities.
The Monte Carlo calculations confirm the expected $\alpha$-independence of these ratios, their constant values being consistent with the grand canonical baseline.

\subsection{Third order cumulants of conserved charges}

Let us consider now third order cumulants.
First we analyze the so-called skewness ratio $\kappa_3^X / \kappa_2^X$ for $X = B, Q, S$.
The skewness is a non-Gaussian fluctuation measure that characterizes the asymmetry of a distribution around the mean value.
The signs of the skewness of QCD conserved charges are thought to be sensitive probes of the QCD phase structure~\cite{Asakawa:2009aj}.
The SAM predicts that the skewness $\kappa_3^X / \kappa_2^X$ scaled by $(1-2\alpha)$ is independent of acceptance and coincides with the skewness $\chi_3^X / \chi_2^X$ evaluated in the grand canonical ensemble.

\begin{figure}[t]
  \centering
  \includegraphics[width=.49\textwidth]{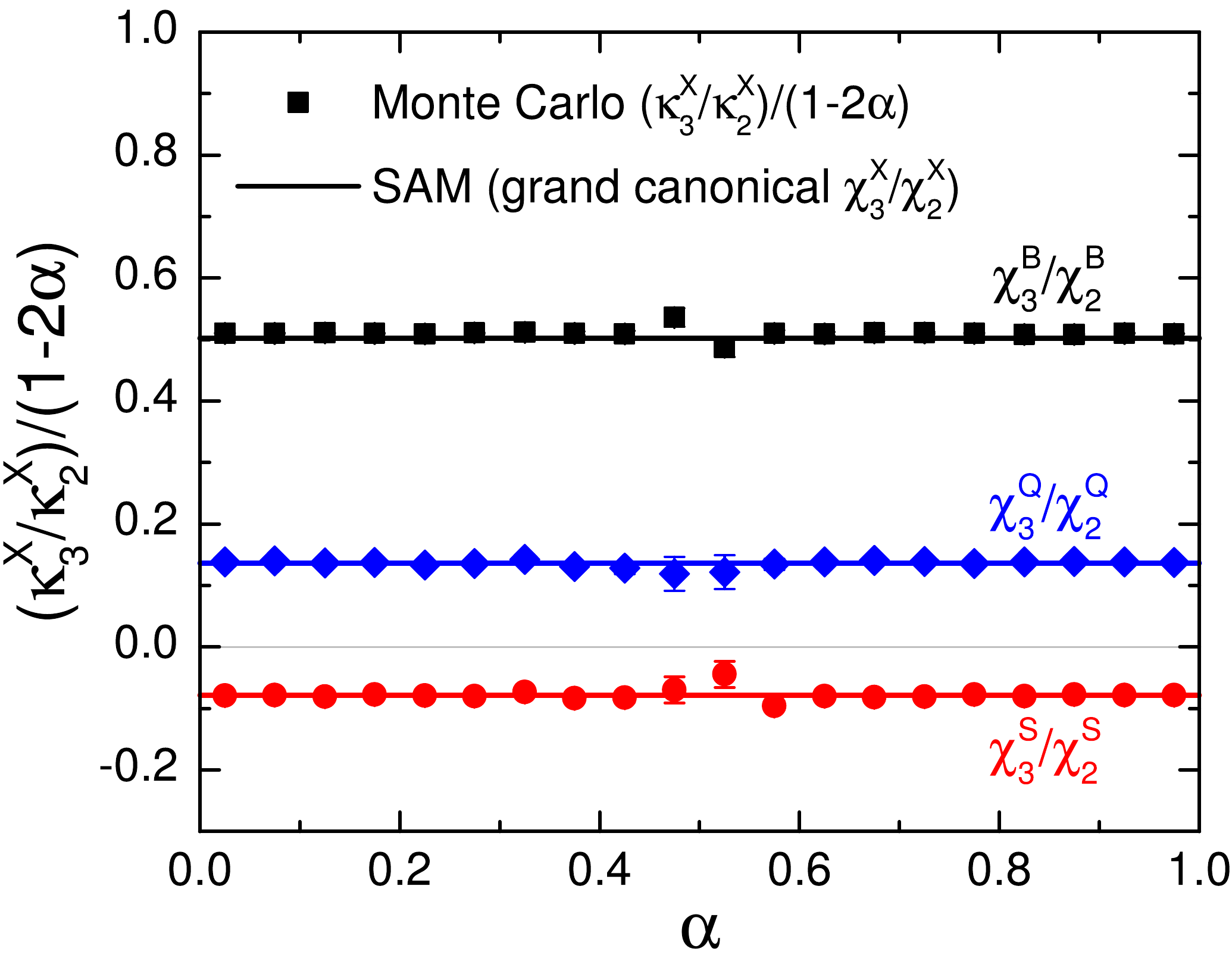}
  \includegraphics[width=.49\textwidth]{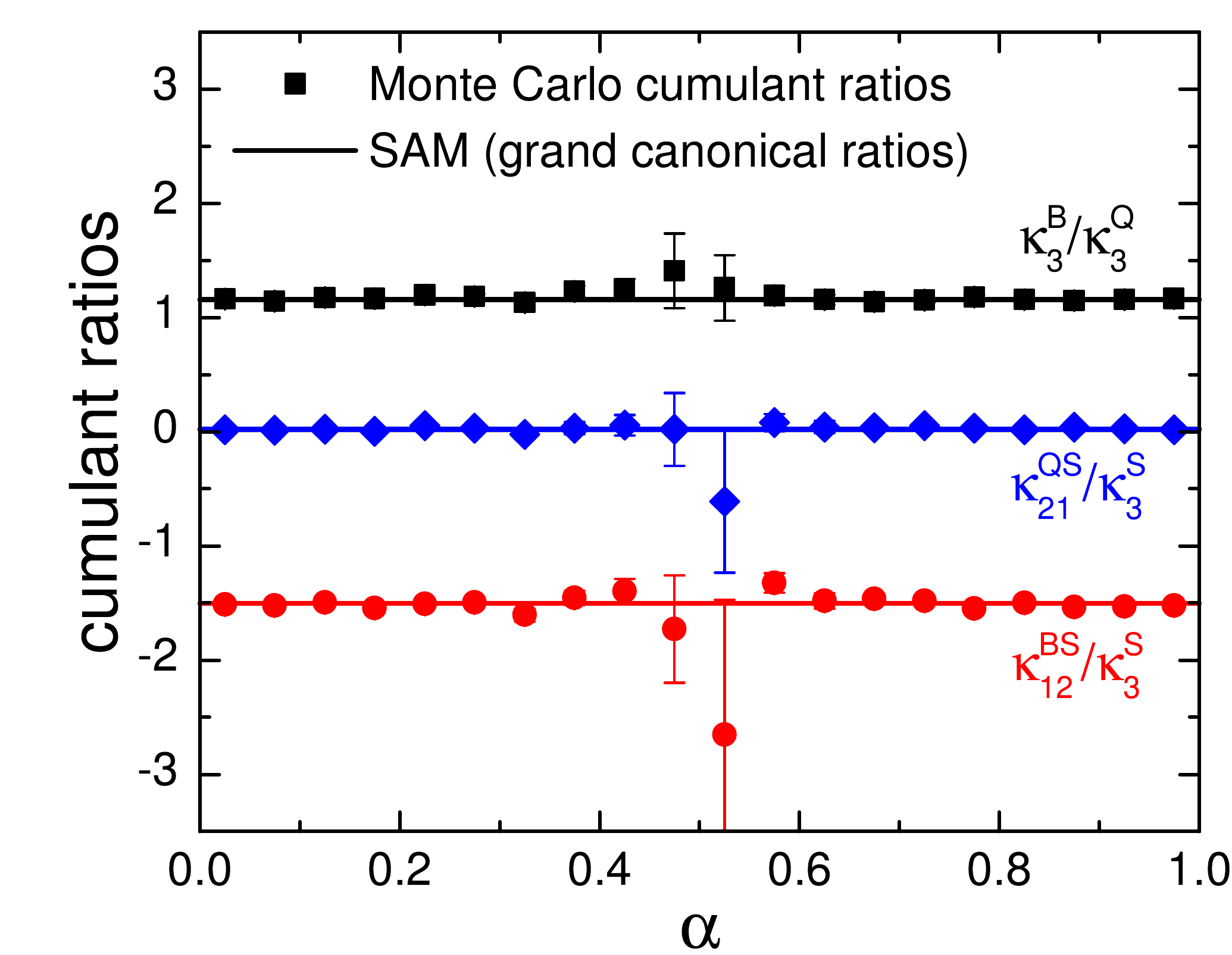}
  \caption{
  Dependence of combinations of third and second order cumulants of conserved charges on the acceptance fraction $\alpha$, as calculated in the hadron resonance gas model using canonical ensemble Monte Carlo sampler~(symbols) and analytically in the framework of the subensemble acceptance method~(lines).
  The Monte Carlo sample is the same as in Fig.~\ref{fig:HRGchi2}.
  \emph{Left panel:} Skewness cumulant ratios $\kappa_3/\kappa_2$ of net-baryon~(black),
  net-charge~(blue), and net-strangeness~(red) numbers scaled by the global charge conservation correction factor $(1-2\alpha)$.  
  \emph{Right panel:} Ratio of third order conserved charge cumulant ratios $\kappa_{3}^{B} / \kappa_3^Q$~(black), $\kappa_{21}^{QS} / \kappa_3^S$~(blue), and $\kappa_{12}^{BS} / \kappa_3^S$~(red).
  Monte Carlo third order cumulants in HRG, diagonal (left) and off-diagonal (right).
  }
  \label{fig:HRGchi3}
\end{figure}

The left panel of Fig.~\ref{fig:HRGchi3} depicts 
the ratios $(\kappa_3^X / \kappa_2^X) / (1-2\alpha)$ for baryon number, electric charge, and strangeness evaluated using the Monte Carlo event generator.
The Monte Carlo results are consistent with $(\kappa_3^X / \kappa_2^X) / (1-2\alpha)$ being
independent on $\alpha$  and coincide with the grand canonical $(\chi_3^X / \chi_2^X)$ susceptibility ratios.

In addition, the SAM predicts that any ratio of two third order
cumulants of conserved charges is insensitive to global conservation laws. This follows from
Eq.~\eqref{eq:kappa3final}.
The right panel of Fig.~\ref{fig:HRGchi3} depicts the $\alpha$-dependence of cumulant ratios $\kappa_{3}^{B} / \kappa_3^Q$, $\kappa_{21}^{QS} / \kappa_3^S$, and $\kappa_{12}^{BS} / \kappa_3^S$, as calculated within the Monte Carlo event generator~(symbols) and within the SAM using the grand canonical susceptiblities.
The Monte Carlo calculations are consistent with the $\alpha$-independence of all these ratios, and in agreement with the grand canonical baseline, as predicted by the SAM.
The statistical errors in the Monte Carlo calculations become large in the vicinity of $\alpha = 1/2$, as clearly seen in Fig.~\ref{fig:HRGchi3}.
This is a consequence of the fact that third order cumulants vanish at $\alpha = 1/2$, as follows from Eq.~\eqref{eq:kappa3final}.
A ratio of third order cumulants in the vicinity of $\alpha = 1/2$ corresponds to a ratio of two small numbers, hence the large statistical uncertainties.
As a consequence, it would be advisable to perform experimental analysis of third order cumulants in acceptances away from $\alpha = 1/2$.

\subsection{Fourth order cumulants of conserved charges}

We turn now to fourth order cumulants.
An interesting new aspect here is that fourth order cumulants are determined not only by the corresponding fourth order GCE susceptibilities but also by second and third order mixed susceptibilities, as seen from Eqs.~\eqref{eq:kappa4final}~[see also Eq.~\eqref{eq:kappa4Bthree}].
The HRG model analysis allows to estimate the importance of these mixed susceptibilities with regard to the behavior of fourth order cumulants in a finite acceptance. 

We shall analyze here the fourth-to-second order ratios $\kappa_4^X / \kappa_2^X$ of diagonal cumulants for
$X = B, Q, S$ -- the so-called kurtosis of a conserved charge distribution.
The kurtosis of the baryon number in a subvolume $V_1 = \alpha V$ is evaluated using Eqs.~\eqref{eq:kappa4Bthree} and \eqref{eq:kappa2B}:
\eq{\label{eq:kurtBQS}
\frac{\kappa_4^B}{\kappa_2^B} & =  \, (1-3\alpha \beta) \, \frac{\chi_4^B}{\chi_2^B} - \frac{3 \alpha \beta}{\chi_2^B \, D[\hat{\chi}_2]} \nonumber \\
& \quad \times \left\{  (\chi_3^B)^2 [\chi_2^Q \chi_2^S - (\chi_{11}^{QS})^2] + (\chi_{21}^{BQ})^2 [\chi_2^B \chi_2^S - (\chi_{11}^{BS})^2] + (\chi_{21}^{BS})^2 [\chi_2^B \chi_2^Q - (\chi_{11}^{BQ})^2]  \right.
\nonumber \\
& \quad \left. 
- 2 \chi_3^B \chi_{21}^{BQ} (\chi_2^S \chi_{11}^{BQ} - \chi_{11}^{BS} \chi_{11}^{QS})
- 2 \chi_3^B \chi_{21}^{BS} (\chi_2^Q \chi_{11}^{BS} - \chi_{11}^{BQ} \chi_{11}^{QS})
\right\}~.
}
The kurtosis of electric charge $Q$ and strangeness $S$ distributions are given by similar expressions, which can be explicitly derived from the general formula~\eqref{eq:kappa4final}.
In fact, these expressions can also be obtained from Eq.~\eqref{eq:kurtBQS} by cyclic permutations
in $(B,Q,S)$.

\begin{figure}[t]
  \centering
  \includegraphics[width=.49\textwidth]{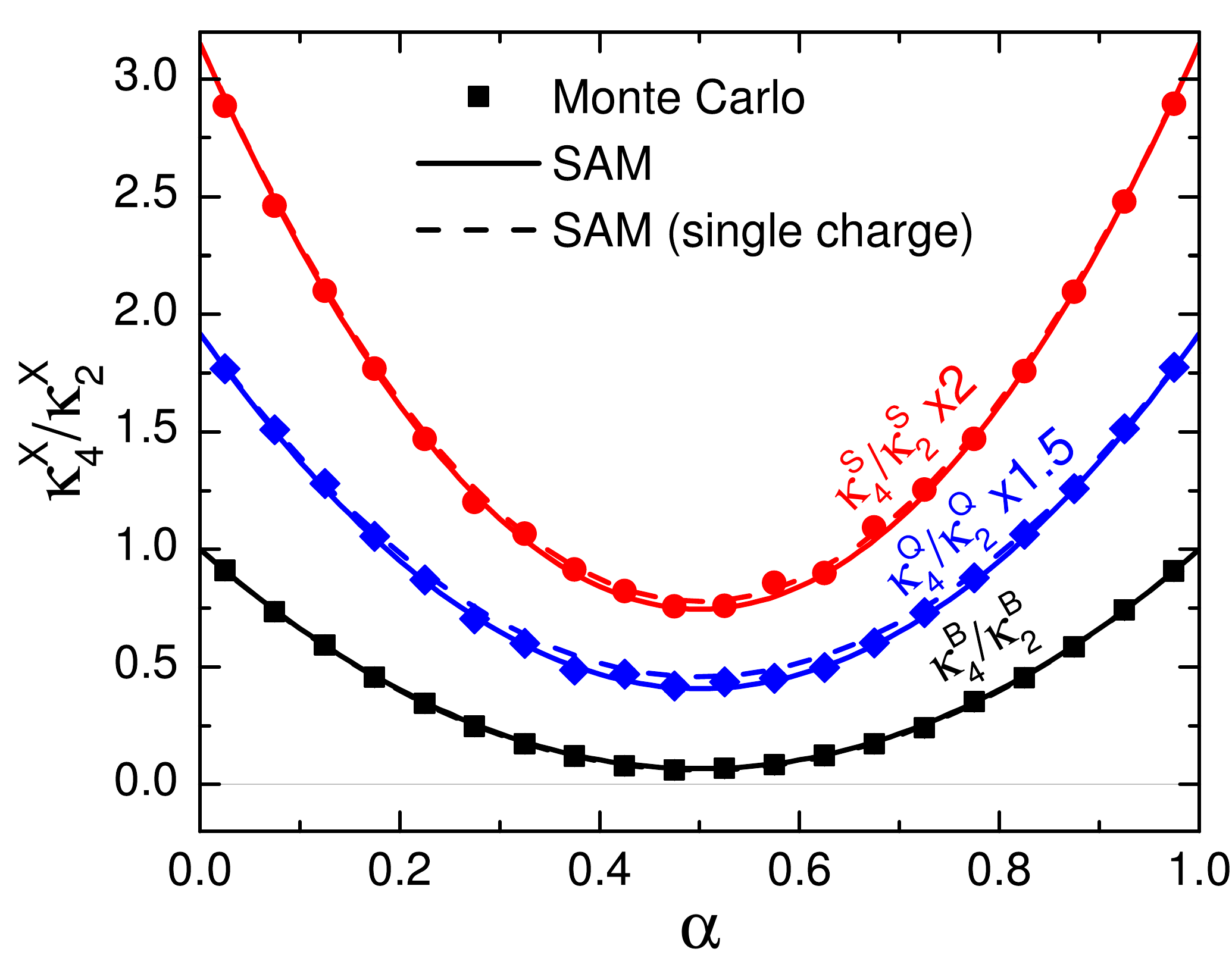}
  \caption{
  Dependence of kurtosis cumulant ratios $\kappa_4/\kappa_2$ of of net-baryon~(black),
  net-charge~(blue), and net-strangeness~(red) numbers on the acceptance fraction $\alpha$, as
  calculated in the hadron resonance gas model using the canonical ensemble Monte Carlo sampler with $10^8$ events~(symbols) and analytically in the framework of the subensemble acceptance method~(solid lines).
  The dashed lines depict SAM calculations for a single conserved charge~[Eq.~\eqref{eq:kurtX}].
  The Monte Carlo sample is the same as in Fig.~\ref{fig:HRGchi2}.
  }
  \label{fig:HRGchi4}
\end{figure}

Figure~\ref{fig:HRGchi4} depicts the $\alpha$-dependence of the kurtosis of baryon number~(black symbols), electric charge~(blue symbols), and net strangeness~(red symbols), as obtained from Monte Carlo simulations.
The solid lines in Fig.~\ref{fig:HRGchi4} correspond to the SAM predictions~[Eq.~\eqref{eq:kurtBQS}], these agree with the Monte Carlo results.
To estimate the relevance of cross-correlations between multiple conserved charges on $\kappa_4^X / \kappa_2^X$ it is instructive to compare the results to predictions of the SAM for a single conserved charge.
The ratio $\kappa_4^X / \kappa_2^X$ in this case is obtained by dividing Eq.~\eqref{eq:kappa4Bsingle} by Eq.~\eqref{eq:kappa2B},
\eq{\label{eq:kurtX}
\frac{\kappa_4^X}{\kappa_2^X} = (1-3\alpha \beta) \, \frac{\chi_4^X}{\chi_2^X} - 3 \, \alpha \, \beta \, \left(\frac{\chi_3^X}{\chi_2^X}\right)^2, \qquad X \in (B,Q,S)~,
}
which is depicted in Fig.~\ref{fig:HRGchi4} by dashed lines.
Equation~\eqref{eq:kurtX} deviates from the general result~\eqref{eq:kurtBQS} by no more than a few percent, indicating the behavior of kurtosis of a conserved charge is primarily driven by the exact conservation of that charge, whereas the influence of exact conservation of other conserved charges is subleading.
We also remind the reader that this influence of other conserved charges vanishes completely at $\mu_B = 0$~(LHC energies), where all odd-order susceptibilities is zero.
These observations lead us to conclude that the simplified Eq.~\eqref{eq:kurtX} is rather accurate for practical applications, at least for $\mu_B \lesssim 100$~MeV~($\sqrt{s_{NN}} \gtrsim 40$~GeV).

\subsection{Off-diagonal cumulants involving non-conserved quantities}

Next, we switch  from cumulants of globally conserved quantities to cumulants involving quantities that are not globally conserved. 
This does better reflect the current experimental reality.
To be more specific, we shall consider second order cumulants involving net-proton, net-kaon, and net-charge numbers.
This is in part motivated by recent experimental efforts of the STAR collaboration in measuring these quantities~\cite{Adam:2019xmk}.
While the net electric charge is globally conserved, the net proton and net kaon numbers do fluctuate even in the full acceptance.

As follows from the results of Sec.~\ref{sec:SEnoncons}, a correlation of a non-conserved quantity with a conserved charge is affected by the global conservation laws by the same factor $(1-\alpha)$ as all second order cumulants of conserved charges.
The implication is that measurable ratios such as $\kappa^{pQ}_{11} / \kappa_2^Q$ and $\kappa^{kQ}_{11} / \kappa_2^Q$ are expected to be unaffected by the global conservation laws and coincide with the corresponding ratios $\chi^{pQ}_{11} / \chi_2^Q$ and $\chi^{kQ}_{11} / \chi_2^Q$ of the grand canonical susceptibilities.

The grand canonical susceptibilities can be evaluated in the HRG model.
Extra care should be taken to account for the large feeddown from decays of resonances to final yields of protons and kaons.
The grand canonical fluctuations and correlations of final particle numbers after resonance decays in HRG have been worked out in Refs.~\cite{Begun:2006jf,Begun:2006uu}.
The grand canonical susceptibility describing correlations between two final-state hadron species $i$ and $j$ reads
\eq{\label{eq:chifinHRG}
\chi_{ij}  \equiv \frac{\mean{\Delta N_i \Delta N_j}}{VT^3} 
 = \delta_{ij} \, \chi_i^{\rm hrg} + \sum_R  \mean{n_i \, n_j}_R \, \chi_R^{\rm hrg} ~.
}
Here the index $R$ sums over all resonances. The quantity $\mean{n_i \, n_j}_R$ is an average
product of the number of hadron species $i$ and $j$ which result from decays of resonance $R$. This
quantity takes into account the multinomial nature of resonance decays. $\chi_i^{\rm hrg}$ is defined in Eq.~\eqref{eq:chiHRG}.

The correlator $\chi_{11}^{kQ}$ of net-kaon number with net-charge is evaluated by incorporating contributions from protons and antiprotons as well as of all charged hadrons in the final state:
\eq{\label{eq:kQHRG}
\chi^{kQ}_{11} \equiv \frac{\mean{\Delta (N_{k^+} - N_{k^-} ) \Delta Q}}{VT^3} 
 = \sum_j q_j (\chi_{k^+ j} - \chi_{k^- j})~.
}
Here $k^+$ and $k^-$ corresponds to positively and negatively charged final state kaons,
respectively, while the index $j$ runs over all hadron species with charge $q_{j}$ in the final state, including both particles and antiparticles.
The expression for the net-proton-net-charge correlator $\chi_{11}^{pQ}$ is analogous to Eq.~\eqref{eq:kQHRG}.
Evaluation of the grand canonical susceptibilities in accordance with Eqs.~\eqref{eq:chifinHRG} and \eqref{eq:kQHRG} is readily implemented in the \texttt{FIST} package, as documented in Ref.~\cite{Vovchenko:2019pjl}.

The Monte Carlo procedure requires some extensions to incorporate the resonance decays as well.
This is done in the following way.
First, we mark all the primordial hadrons and resonances as either belonging to subvolume $V_1$ or
subvolume $V_{2}$ by doing the Bernoulli trials, as before.
Then we let all unstable resonances  decay until only the stable decay products are left. 
All decay products stemming from the same primordial resonance are assigned the same subvolume as that primordial resonance.
We generate $10^7$ events where we perform resonance decays in accordance with the above description.
All the cumulants of interest are then computed in the standard way, as a statistical average from final-state hadron distributions.

\begin{figure}[t]
  \centering
  \includegraphics[width=.49\textwidth]{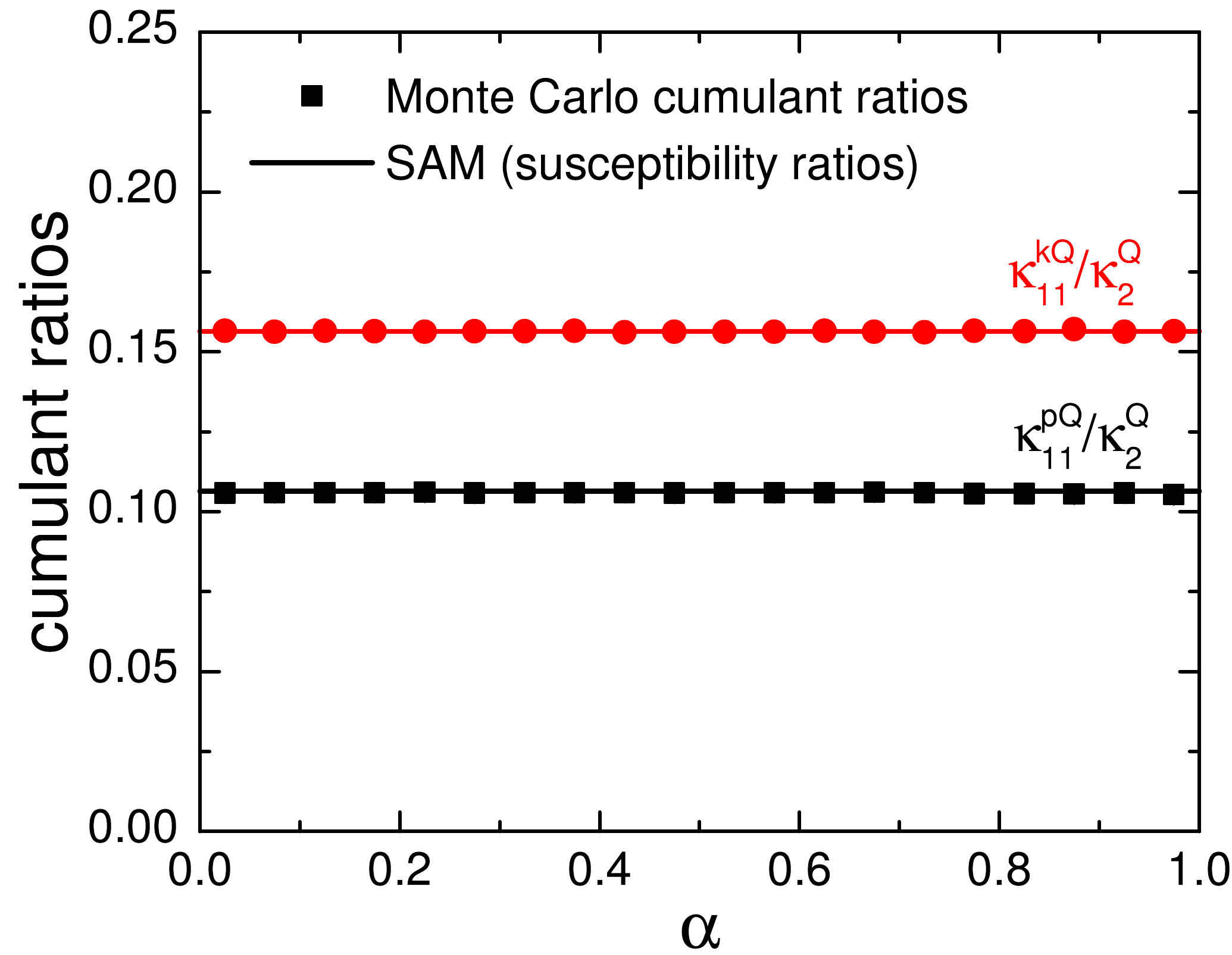}
  \includegraphics[width=.49\textwidth]{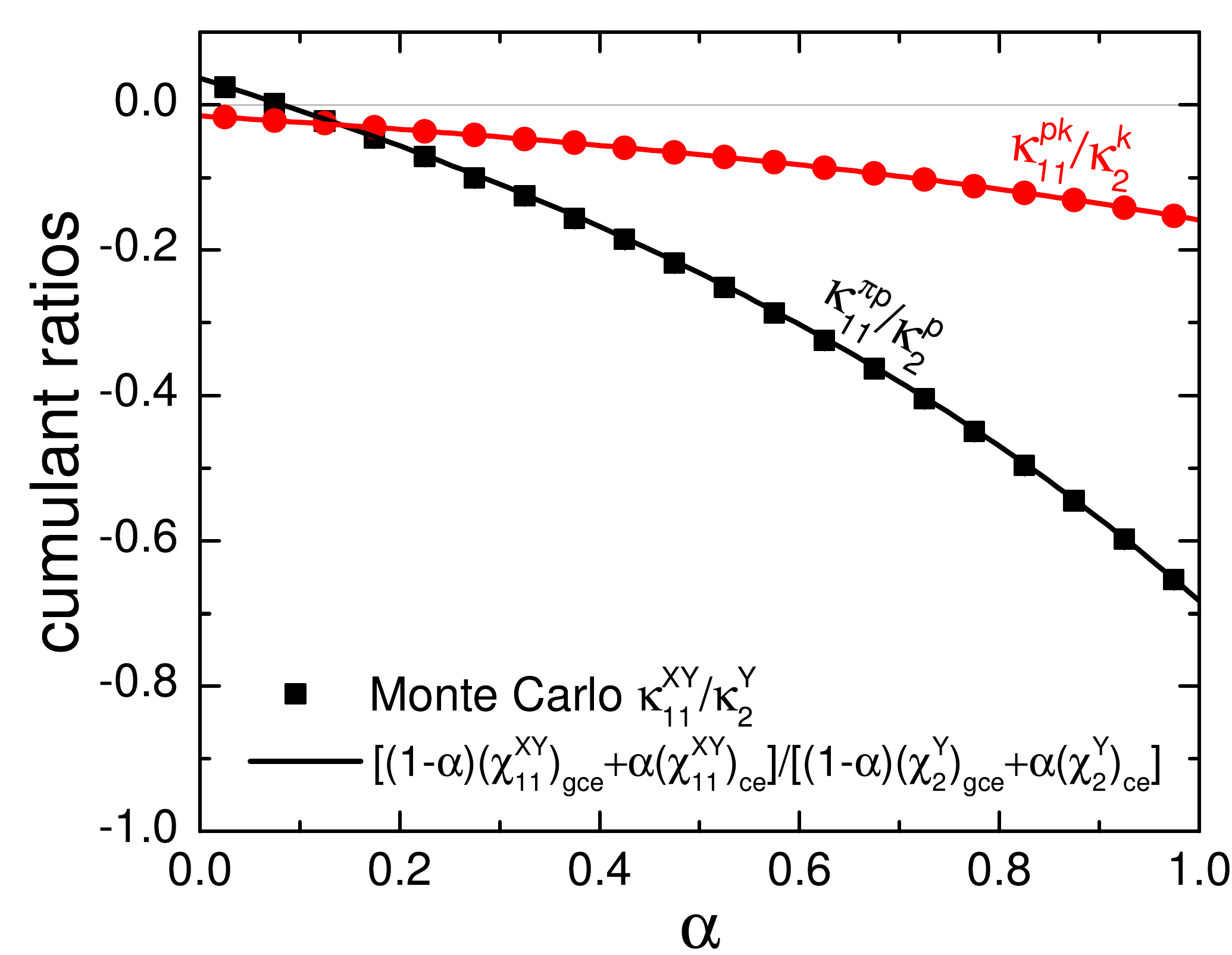}
  \caption{
  Dependence of cumulant ratios $\kappa_{11}^{pQ} / \kappa_2^Q$ and $\kappa_{11}^{kQ} / \kappa_2^Q$~(\emph{left panel}) and $\kappa_{11}^{\pi p} / \kappa_2^p$ and $\kappa_{11}^{pk} / \kappa_2^k$~(\emph{right panel}) on the acceptance $\alpha$, as calculated in the hadron resonance gas model using canonical ensemble Monte Carlo sampler~(symbols) and analytically in the framework of the subensemble acceptance method~(lines).
  Here $p$, $k$, and $\pi$ stand, respectively, for net-proton, net-kaon and net-pion numbers evaluated after resonance decays.
  The Monte Carlo sample contains $10^7$ events and uses the same thermal parameters as in Fig.~\ref{fig:HRGchi2}.
  }
  \label{fig:HRGchi2noncons}
\end{figure}

The left panel of Fig.~\ref{fig:HRGchi2noncons} depicts the Monte Carlo calculation results for cumulant ratios $\kappa^{pQ}_{11} / \kappa_2^Q$ and $\kappa^{kQ}_{11} / \kappa_2^Q$ as functions of acceptance parameter $\alpha$.
The ratios do not exhibit any sensitivity to the value of $\alpha$.
As predicted by the SAM, these quantities coincide with the corresponding ratios $\chi^{pQ}_{11} / \chi_2^Q$ and $\chi^{kQ}_{11} / \chi_2^Q$ of the grand canonical susceptibilities, evaluated through Eqs.~\eqref{eq:chifinHRG} and~\eqref{eq:kQHRG}.
Measurements of such quantities can thus directly reflect intrinsic properties of matter that are characterized by the grand canonical susceptibilities.

Second order cumulant ratios involving non-conserved quantities only, on the other hand, do depend on the size of the subvolume.
Effects of global conservation laws no longer cancel out in such a case.
To illustrate this aspect we show in the right panel of Fig.~\ref{fig:HRGchi2noncons} the $\alpha$-dependence of cumulant ratios $\kappa^{\pi p}_{11} / \kappa_2^p$ and $\kappa^{pk}_{11} / \kappa_2^k$ involving correlations of net proton number with net pion and net kaon numbers, respectively.
These ratios clearly do exhibit $\alpha$ dependence, interpolating between ratios of
grand canonical~($\alpha \to 0$) and canonical~($\alpha \to 1$) susceptibilities. 

The SAM predicts a linear $\alpha$-dependence of net-proton and net-kaon cumulants $\kappa_2^p$ and $\kappa_2^k$~[Eq.~\eqref{eq:kappapp}].
Assuming the same linear $\alpha$-dependence holds also for the correlators $\kappa^{p\pi}_{11}$ and $\kappa^{pk}_{11}$ one can expresses the ratios $\kappa^{\pi p}_{11} / \kappa_2^p$ and $\kappa^{pk}_{11} / \kappa_2^k$ as follows:
\eq{\label{eq:kappappi}
\frac{\kappa^{\pi p}_{11}}{\kappa_2^p}
& =
\frac{(1-\alpha) \chi_{11}^{\pi p} + \alpha \, (\chi_{11}^{\pi p})_{\rm ce}}{(1-\alpha) \chi_{2}^{p} + \alpha \, (\chi_{2}^{p})_{\rm ce}}~, \\
\label{eq:kappapk}
\frac{\kappa^{pk}_{11}}{\kappa_2^k}
& =
\frac{(1-\alpha) \chi_{11}^{pk} + \alpha \, (\chi_{11}^{pk})_{\rm ce}}{(1-\alpha) \chi_{2}^{k} + \alpha \, (\chi_{2}^{k})_{\rm ce}}~.
}
Here the quantities denoted as $(\ldots)_{\rm ce}$ are the susceptibilities evaluated in the
canonical ensemble. The procedure to evaluate such quantities analytically is detailed in
Refs.~\cite{Begun:2006jf,Begun:2006uu}, and is readily available in the \texttt{FIST} package. 
The $\alpha$-dependence of $\kappa^{p\pi}_{11} / \kappa_2^p$ and $\kappa^{pk}_{11} / \kappa_2^k$ evaluated according to Eqs.~\eqref{eq:kappappi} and \eqref{eq:kappapk} is shown in the right panel of Fig.~\ref{fig:HRGchi2noncons} by dashed lines.
These analytic expectations agree with the Monte Carlo results, suggesting that all second order cumulants of non-conserved quantities can be generally described by a linear function in $\alpha$ which interpolates between the grand canonical~($\alpha \to 0$) and canonical~($\alpha \to 1$) limits, as written in the numerator and denominator of Eqs.~\eqref{eq:kappappi} and \eqref{eq:kappapk}.

\subsection{Net-proton and net-$\Lambda$ fluctuations}
\label{sec:netPnetL}

We shall conclude our HRG model analysis by exploring fluctuations of net-proton and net-$\Lambda$ number.
These quantities are more accessible to experimental measurements than net-baryon number fluctuations.
The grand canonical HRG model provides a natural baseline for the second order cumulants of net-proton~(net-$\Lambda$) fluctuations: the net-proton~(net-$\Lambda$) number fluctuations are described by the Skellam distribution, meaning that the second order cumulant simply counts the mean number of protons~($\Lambda$) and antiprotons~($\bar{\Lambda}$):
\eq{
\kappa_2^{\rm Sk}[p - \bar{p}] & = VT^3 \, \chi_2^p = 
\mean{N_p} + \mean{N_{\bar{p}}}~, 
\\
\kappa_2^{\rm Sk}[\Lambda - \bar{\Lambda}] & = VT^3 \, \chi_2^{\Lambda} = 
\mean{N_{\Lambda}} + \mean{N_{\bar{\Lambda}}}~.
}
This baseline is not modified by resonance decays since no decays are known to generate proton-proton~($\Lambda\Lambda$) or proton-antiproton~($\Lambda\bar{\Lambda}$) correlations.

Naturally, both the net proton and net $\Lambda$ fluctuations are affected by the exact global conservation of 
charges.
In particular, the effect of baryon number conservation on net-proton fluctuations has extensively
been studied in the literature within the HRG model with a canonical treatment of baryon number~\cite{Bzdak:2012an,Braun-Munzinger:2016yjz}.
If the acceptance for particles and antiparticles is uniform, the effect of global baryon number conservation on net-proton fluctuations is the following:
\eq{\label{eq:kappa2pprev}
\frac{\kappa_2[p - \bar{p}]}{\kappa_2^{\rm Sk}[p - \bar{p}]} = 1 - \alpha_B^p~, \qquad \qquad \alpha_B^p = \frac{\mean{N_p^{\rm acc}} + \mean{N_{\bar{p}}^{\rm acc}}}{\mean{N_B^{\rm 4\pi}} + \mean{N_{\bar{B}}^{\rm 4\pi}}}~.
}
Here $\mean{N_{p(\bar{p})}^{\rm acc}}$ is the mean number of (anti)protons within the acceptance where net-proton fluctuations are measured and $\mean{N_{B(\bar{B})}^{\rm 4\pi}}$ is the mean number of (anti)baryons in the full space.
The expression for net-$\Lambda$ fluctuations is analogous.
Note that $\alpha_B^p$ can also be defined as $\alpha_B^p = \mean{N_p^{\rm acc}} / \mean{N_B^{\rm 4\pi}}$~\cite{Bzdak:2012an,Braun-Munzinger:2016yjz} if the acceptance factor is the same for protons and antiprotons. 
This is the case at the LHC due to particle-antiparticle symmetry at $\mu_B = 0$, as well as at low collision energies where the production of antibaryons can be neglected.
In a general case, however, differences between the two definitions do appear because of non-zero values of $\mu_Q$ and $\mu_S$.
We checked that these differences do not exceed 2\% at all collision energies in the HRG model,
therefore, in practice either of the two definitions may  be adopted.

Expression~\eqref{eq:kappa2pprev} has been used by the ALICE collaboration to interpret their measurements of the behavior of net-proton fluctuations in Pb-Pb collisions at the LHC as primarily driven by an effect of global baryon number conservation~\cite{Acharya:2019izy}.
At the same time, fluctuations can also be influenced by exact conservation of electric charge and strangeness, in particular since protons do carry electric charge and $\Lambda$'s do carry strangeness.
This point has been made in a recent STAR paper on net-$\Lambda$ fluctuations~\cite{Adam:2020kzk}, where an ad hoc relation $\kappa_2 [\Lambda - \bar{\Lambda}] / \kappa_2^{\rm NBD}[\Lambda - \bar{\Lambda}] = 1 - (\alpha_B^\Lambda + \alpha_S^\Lambda)$ has been used to estimate the simultaneous effect of net baryon and net strangeness conservation.
In the STAR paper~\cite{Adam:2020kzk} the baseline $\kappa_2^{\rm NBD}[\Lambda - \bar{\Lambda}]$ corresponds to a negative binomial distribution, which is similar although slightly different than the Skellam baseline~$\kappa_2^{\rm Sk}[\Lambda - \bar{\Lambda}]$ discussed here.

Here we shall employ the SAM to rigorously study the effect of multiple conserved charges on net-proton and net-$\Lambda$ fluctuations.
Our starting point is Eq.~\eqref{eq:kappapp} which describes the second order cumulant of an arbitrary non-conserved quantity within a subvolume for an arbitrary equation of state.
We rewrite Eq.~\eqref{eq:kappapp} here for net-protons in a HRG model:
\eq{
\kappa_2[p_1 - \bar{p}_1] & = 
\alpha VT^3 \left[ (1-\alpha) \chi_2^p +\alpha \frac{\det{\breve{\chi}^{\rm hrg}}}{\det{\chi^{\rm hrg}}} \right].
}
Normalizing the net proton cumulant by the Skellam baseline, $\kappa_2^{\rm Sk}[p_1 - \bar{p}_1] = \alpha V T^3 \, \chi_2^p$, we get
\eq{\label{eq:kappa2poverSk}
\frac{\kappa_2[p_1 - \bar{p}_1]}{\kappa_2^{\rm Sk}[p_1 - \bar{p}_1]} = 1 - \alpha + \frac{\alpha}{\chi_2^p} \frac{\det{\breve{\chi}^{\rm hrg}}}{\det{\chi^{\rm hrg}}}~.
}

Let us 
first consider
a single conserved charge -- the baryon number $B$.
In this case $\det{\chi^{\rm hrg}} = \chi_2^B$ and $\det{\breve{\chi}^{\rm hrg}} = \chi_2^B \chi_2^p - (\chi_{11}^{pB})^2$.
Furthermore, in the HRG model one has $\chi_{11}^{pB} = \chi_2^p$.
Inserting these relations into Eq.~\eqref{eq:kappa2poverSk} one obtains
\eq{
\frac{\kappa_2^B[p_1 - \bar{p}_1]}{\kappa_2^{\rm Sk}[p_1 - \bar{p}_1]} = 1 - \alpha \frac{\chi_2^p}{\chi_2^B}~.
}
In the HRG model $\chi_2^p$ and $\chi_2^B$ are proportional to the {\em total} number of protons
plus antiprotons and to the {\em total} number of baryons plus antibaryons, respectively, with the
same proportionality factor, so that
$\chi_2^p / \chi_2^B = (\mean{N_p^{4\pi}} + \mean{N_{\bar{p}}^{4\pi}}) / (\mean{N_B^{4\pi}} + \mean{N_{\bar{B}}^{4\pi}})$.
The subvolume fraction $\alpha$ is proportional to the ratio of mean number of protons plus antiprotons in acceptance to the mean number of protons plus antiprotons in the full space, $\alpha = (\mean{N_p^{\rm acc}} + \mean{N_{\bar{p}}^{\rm acc}})/(\mean{N_p^{4\pi}} + \mean{N_{\bar{p}}^{4\pi}})$.
Therefore, $\alpha \chi_2^p / \chi_2^B = (\mean{N_p^{\rm acc}} + \mean{N_{\bar{p}}^{\rm acc}}) / (\mean{N_B^{4\pi}} + \mean{N_{\bar{B}}^{4\pi}}) = \alpha_B^p$ and
\eq{\label{eq:kappa2pB}
\frac{\kappa_2^B[p_1 - \bar{p}_1]}{\kappa_2^{\rm Sk}[p_1 - \bar{p}_1]} = 1 - \alpha_B^p~,
}
in agreement with Eq.~\eqref{eq:kappa2pprev}.
One gets an analogous relation for net-$\Lambda$ fluctuations using the same logic.

Let us now consider the exact conservation of electric charge $Q$ in addition to baryon number $B$.
In this case the susceptibilities involving $Q$ contribute to $\det{\chi^{\rm hrg}}$ and 
$\det{\breve{\chi}^{\rm hrg}}$ in the r.h.s. of Eq.~\eqref{eq:kappa2poverSk}.
To simplify the resulting expression we shall make use of the fact that resonance decays generate negligibly small proton-charge correlations in addition to proton self-correlation, implying that $\chi_{11}^{pQ} \approx \chi_2^p$ holds to a large precision.\footnote{We note that $\Delta(1232)^{++} \to p + \pi^+$ decays generate an excess of $\chi_{11}^{pQ}$ over $\chi_2^p$.
On the other hand, decays like $\Delta(1232)^{0} \to p + \pi^-$ lead to a reduction of $\chi_{11}^{pQ}$ relative to $\chi_2^p$.
Our HRG model calculations reveal only percent level deviations of $\chi_{11}^{pQ}$ from $\chi_2^p$ after all resonance decays are accounted for.
}
After some algebra, one obtains
\eq{\label{eq:kappapBQ}
\frac{\kappa_2^{BQ}[p_1 - \bar{p}_1]}{\kappa_2^{\rm Sk}[p_1 - \bar{p}_1]} = 1 - (\alpha_B^p + \alpha_Q^p) \, \frac{1- \ddfrac{2 \, \chi_{11}^{BQ}}{\chi_2^B + \chi_2^Q} }{1 - \ddfrac{(\chi_{11}^{BQ})^2}{\chi_2^B  \chi_2^Q} }~, \qquad \qquad \chi_{11}^{pQ} \approx \chi_2^p~.
}
Here 
\eq{\label{eq:alphapQ}
\alpha_Q^p = \alpha \frac{\chi_2^p}{\chi_2^Q} \approx \frac{\mean{N_p^{\rm acc}}}{\mean{N_{\rm ch, prim}^{4\pi}}}~
}
and $\mean{N_{\rm ch, prim}^{4\pi}}$ corresponds to the charged particle multiplicity at the chemical
  freeze-out stage, i.e. {\em before } resonance decays.
If contributions of multi-charged hadrons are small, the approximate relation in Eq.~\eqref{eq:alphapQ} becomes exact.
Note that the final charged multiplicity $\mean{N_{\rm ch, fin}^{4\pi}}$ measured in the experiment can be considerably larger than $\mean{N_{\rm ch, prim}^{4\pi}}$, due to decays of neutral resonances into charged particles, e.g. the $\rho^0 \to \pi^+ + \pi^-$ decay.
This effect is significant, and our HRG calculations suggest that $\mean{N_{\rm ch, fin}^{4\pi}}$ can be up to a factor two larger than $\mean{N_{\rm ch, prim}^{4\pi}}$ at RHIC and LHC energies.
For this reason a reliable estimation of $\alpha_Q^p$ can be challenging.
Taking $\mean{N_{\rm ch, fin}^{4\pi}}$ in place of $\mean{N_{\rm ch, prim}^{4\pi}}$ in Eq.~\eqref{eq:alphapQ} will underestimate the value of $\alpha_Q^p$.

For the electrically neutral $\Lambda$-hyperon it makes sense to consider global conservation of strangeness instead of electric charge.
The simultaneous effect of baryon number and strangeness conservation on net-$\Lambda$ fluctuations is the following:
\eq{\label{eq:kappaLBS}
\frac{\kappa_2^{BS}[\Lambda_1 - \bar{\Lambda}_1]}{\kappa_2^{\rm Sk}[\Lambda_1 - \bar{\Lambda}_1]} & = 1 - (\alpha_B^\Lambda + \alpha_S^\Lambda) \, \frac{1 + \ddfrac{2 \, \chi_{11}^{BS}}{\chi_2^B + \chi_2^S} }{1 - \ddfrac{(\chi_{11}^{BS})^2}{\chi_2^B  \chi_2^S} }~, \qquad \qquad \chi_{11}^{\Lambda S} \approx -\chi_2^\Lambda~, \\
\label{eq:alphaLS}
\alpha_S^\Lambda & = \alpha \frac{\chi_2^\Lambda}{\chi_2^S} \approx \frac{\mean{N_\Lambda^{\rm acc}}}{\mean{N_{S,\rm prim}^{4\pi}}}~.
}
Here $\mean{N_{S,\rm prim}^{4\pi}}$ counts the total number of strange hadrons.
As in the case of net-charge in Eq.~\eqref{eq:alphapQ}, $\mean{N_{S,\rm prim}^{4\pi}}$ corresponds to  hadrons before resonance decays. 
The decays, however, lead only to a small distortion, primarily through a $\phi \to K^+ K^-$ decay, thus $\mean{N_{S,\rm prim}^{4\pi}} \approx \mean{N_{S,\rm fin}^{4\pi}}$ to a very
good approximation.

Equations~\eqref{eq:kappapBQ} and~\eqref{eq:kappaLBS} allow to establish the theoretical basis for the ad hoc relation proposed in Ref.~\cite{Adam:2020kzk} to take into account simultaneous global conservation of two conserved charges.
That relation works for net-proton and net-$\Lambda$ fluctuations if, respectively, baryon-electric and baryon-strangeness correlations in the equation of state can be neglected, i.e.
\eq{\label{eq:k2pBQappr}
\frac{\kappa_2^{BQ}[p_1 - \bar{p}_1]}{\kappa_2^{\rm Sk}[p_1 - \bar{p}_1]} 
& \approx 1 - (\alpha_B^p + \alpha_Q^p)~, \qquad \qquad \text{if }  \chi_{11}^{BQ} \ll \chi_2^B, \chi_2^Q, \\
\label{eq:k2LBSappr}
\frac{\kappa_2^{BS}[\Lambda_1 - \bar{\Lambda}_1]}{\kappa_2^{\rm Sk}[\Lambda_1 - \bar{\Lambda}_1]} 
& \approx 1 - (\alpha_B^\Lambda + \alpha_S^\Lambda)~, \qquad \qquad \text{if }   \chi_{11}^{BS} \ll \chi_2^B, \chi_2^S.
}
Below we present explicit calculations to establish the accuracy of these relations.

\begin{figure}[t]
  \centering
  \includegraphics[width=.49\textwidth]{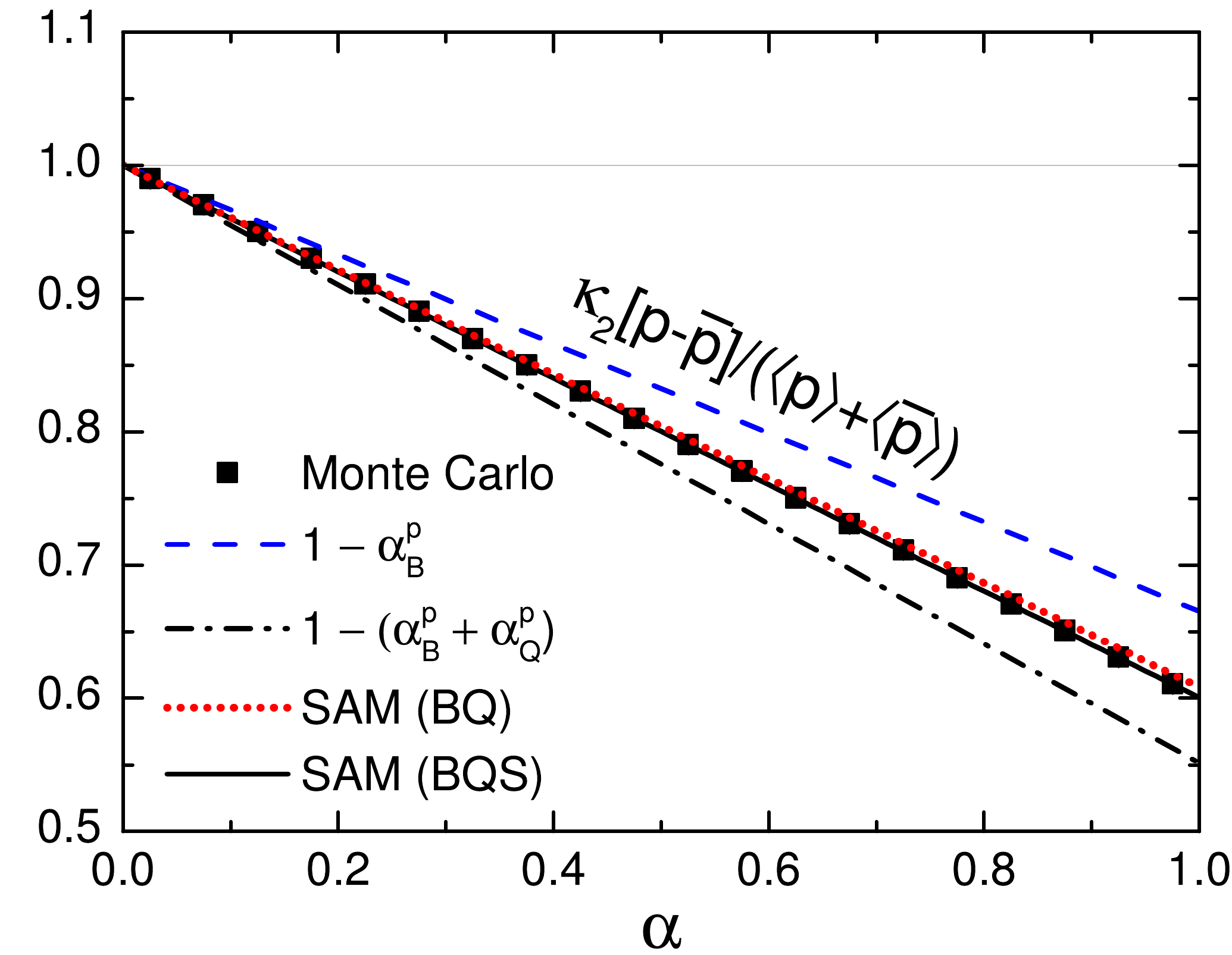}
  \includegraphics[width=.49\textwidth]{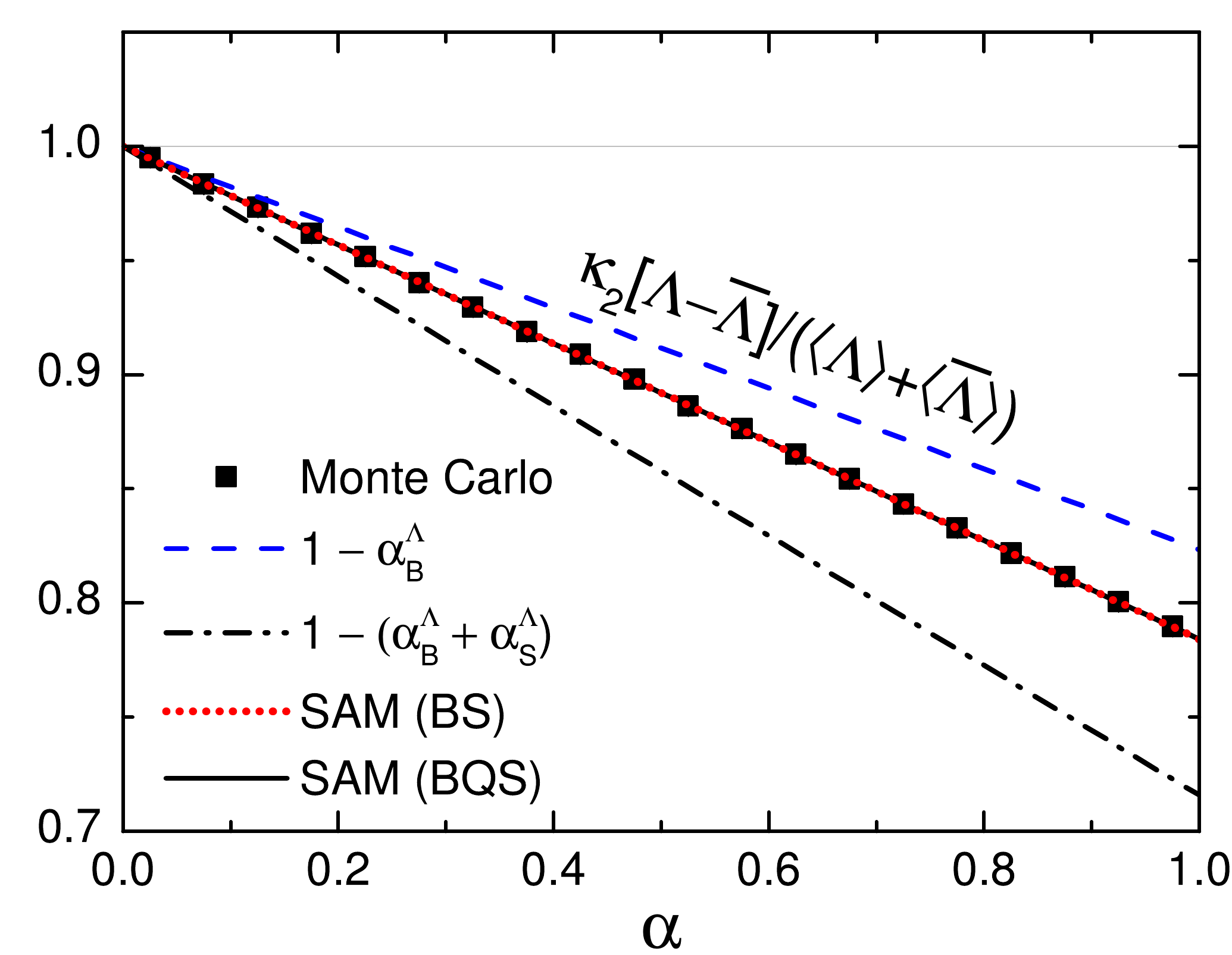}
  \caption{
  Acceptance $\alpha$ dependence of net-particle variance normalized by Skellam distribution baseline for net-proton~(\emph{left panel}) and net-$\Lambda$~(\emph{right panel}) fluctuations, as calculated in the hadron resonance gas model using canonical ensemble Monte Carlo sampler~(symbols) and analytically in the framework of the subensemble acceptance method~(SAM) with three conserved charges~(solid black lines), two conserved charges~(dotted red lines), one conserved charge~(dashed blue lines), as well as using approximate relations~\eqref{eq:k2pBQappr} and \eqref{eq:k2LBSappr} for two conserved charges~(dash-dotted black lines).
  The Monte Carlo sample is the same as in Fig.~\ref{fig:HRGchi2noncons}.
  }
  \label{fig:HRGchi2protLamb}
\end{figure}

Figure~\ref{fig:HRGchi2protLamb} depicts the $\alpha$-dependence of the second order net-proton~(left panel) and net-$\Lambda$~(right panel) cumulants scaled by the grand canonical Skellam distribution baselines.
Calculations include feeddown from all strong and electromagnetic decays.
The symbols depict the results of Monte Carlo sampling.
The Monte Carlo calculations agree with the exact SAM calculation 
in a presence of three
canonically conserved charges~[Eq.~\eqref{eq:kappa2poverSk}], shown in
Fig.~\ref{fig:HRGchi2protLamb} by the solid lines. 

The dotted lines in Fig.~\ref{fig:HRGchi2protLamb} depict the SAM results for two conserved charges,
($B,Q$)
in case of net protons and 
($B,S$)
in case
of net-$\Lambda$. 
Here {\em no} assumptions were made about the approximate equality of the diagonal and off-diagonal second order cumulants as in Eqs.~(\ref{eq:kappapBQ}),(\ref{eq:kappaLBS}). 
These show only negligible deviations from the full $BQS$-canonical ensemble, indicating that the exact conservation of net strangeness has a negligible effect on net-proton fluctuations while the  exact conservation of electric charge has a negligible effect on net-$\Lambda$ fluctuations.
On the other hand, the approximate relations~\eqref{eq:k2pBQappr} and~\eqref{eq:k2LBSappr}, shown in Fig.~\ref{fig:HRGchi2protLamb} by the dash-dotted lines, reveal sizable deviations from the full results.
These relations overestimate the global charge conservation effects, especially at larger values of $\alpha$.
The main reason is that the approximations that go into Eqs.~\eqref{eq:k2pBQappr} and~\eqref{eq:k2LBSappr}, namely the assumed smallness of baryon-electric and baryon-strangeness correlators, are not very accurate at the chemical freeze-out, as follows from the HRG model calculations.\footnote{We used Eqs.~\eqref{eq:alphapQ} and~\eqref{eq:alphaLS} involving primordial numbers of charged and strange hadrons in this calculation.
The result for net-$\Lambda$ would change very little if we used $\mean{N_{S,\rm fin}^{4\pi}}$ instead of $\mean{N_{S,\rm prim}^{4\pi}}$.
Had we used $\mean{N_{\rm ch,\rm fin}^{4\pi}}$ instead of $\mean{N_{\rm ch,\rm prim}^{4\pi}}$, however, it would move the black dash-dotted line in Fig.~\ref{fig:HRGchi2protLamb} up closer to the Monte Carlo data as the fact that $\mean{N_{\rm ch,\rm fin}^{4\pi}} > \mean{N_{\rm ch,\rm prim}^{4\pi}}$ implies a decrease of $\alpha_Q^p$.
}
The approximations $\chi_{11}^{pQ} \approx \chi_2^p$~[Eq.~\eqref{eq:kappapBQ}] and $\chi_{11}^{\Lambda S} \approx \chi_{2}^{\Lambda}$~[Eq.~\eqref{eq:kappaLBS}], on the other hand, are found to be very accurate, namely within 1 and 2\% relative error, respectively.

Finally, the dashed lines in Fig.~\ref{fig:HRGchi2protLamb} depict the net-proton and net-$\Lambda$ fluctuations in the presence of only single conserved charge -- the baryon number~[Eq.~\eqref{eq:kappa2pB}].
Such a calculation is relatively close to the full result, although it does systematically underestimate the overall effect of three conserved charges.

We conclude that a quantitative HRG model analysis of net-proton and net-$\Lambda$ fluctuations requires, in addition to baryon number, a canonical treatment of, respectively, electric charge and strangeness.
This may be even more relevant for the higher-order fluctuations, which are expected to be more sensitive to exact conservation of multiple charges.
Furthermore, in the low-energy limit of heavy-ion collisions, where the production of pions and antibaryons can be neglected, 
the net-proton number coincides with the net electric charge\footnote{Note that one has to take into account contributions from both the unbound protons and protons bound in light nuclei.}, 
thus the influence of global conservation laws on net-proton number will be entirely driven by electric charge conservation.
From a practical point of view, the approximate relations~\eqref{eq:kappa2pB} and~\eqref{eq:k2pBQappr},\eqref{eq:k2LBSappr} can be used to estimate the magnitude of the global charge conservation effects, as they are found to bracket the true values of net-proton and net-$\Lambda$ cumulants in an HRG model calculation.

\section{Discussion and conclusions}
\label{sec:concl}

In this work we extended the subensemble acceptance method, originally formulated in Ref.~\cite{Vovchenko:2020tsr} for a single conserved charge, to the case of multiple conserved charges. 
This allowed us to express cumulants of conserved charge distributions measured in a subvolume of a thermal system with a globally conserved charge in terms of the grand canonical susceptibilities for any equation of state. 
Explicit expressions have been provided for all diagonal and off-diagonal cumulants up to the sixth order, the formalism permits iterative computation of higher-order cumulants as well, if desired.

Among the many results that follow from our formalism we would like to highlight the following two observations:

\begin{itemize}

    \item For cumulants up to third order, the effect of global conservation laws and equation of state factorizes into a product of cumulants of the binomial and grand-canonical distributions~[see Eqs.~\eqref{eq:kappa1final}-\eqref{eq:kappa3final}].
    As a consequence, the global conservation effects cancel out in any ratio of two second order cumulants and in any ratio of two third order cumulants, these quantities simply reduce to ratios of the corresponding grand canonical susceptibilities. 
    We verified this statement explicitly using Monte Carlo sampling of the canonical hadron resonance gas model shown in Figs.~\ref{fig:HRGchi2} and~\ref{fig:HRGchi3}.
    We thus argue that such quantities are particularly suitable for experimental measurements, as they allow to eliminate the dependence of results on a relatively difficult-to-constrain value of the acceptance parameter $\alpha$.
    
    \item The kurtosis of a conserved charge distribution is affected by conservation laws involving other conserved charges, see e.g.~Eq.~\eqref{eq:kappa4Bthree}.
    Our HRG model analysis suggests that this effect is small in heavy-ion collisions at $\sqrt{s}_{\rm NN} \gtrsim 40$~GeV~(see Fig.~\ref{fig:HRGchi4}), implying that the kurtosis of a conserved charge is mainly affected by the exact conservation of that charge, while the conservation laws involving other conserved charges have a small effect at those energies. 
    The discussion of the low collision energy limit in Eq.~\eqref{eq:kappa4Blowenergy} indicates that effects of multiple conserved charges become small there as well. 
    As this result is somewhat counter-intuitive, it indicates that analysis at lower collision energies should be done with care.
    Note that the effects of multiple conserved charges can be more substantial in cumulants of conserved charges of higher order, or already in second cumulants of non-conserved quantities like net-proton or net-$\Lambda$ number, as suggested by our HRG model calculations in Fig.~\ref{fig:HRGchi2protLamb}.

\end{itemize}

We summarize our other findings as follows:

\begin{itemize}
    
    \item From an experimental point of view, it is challenging to measure fluctuations of conserved charges other than electric charge $Q$. In Sec.~\ref{sec:SEnoncons} we have extended the SAM to incorporate fluctuations and correlations involving non-conserved quantities, such as net-proton or net-kaon number.
    Our main result here is Eq.~\eqref{eq:pQfinal}: a correlator of non-conserved quantity, such as net-proton number, with a conserved charge, such as $Q$, is affected by global conservation laws by the same factor as any second order cumulant of conserved charges.
    As a consequence, the readily measurable cumulant ratios such as $\kappa_{11}^{pQ} /
    \kappa_2^Q$, $\kappa_{11}^{kQ} / \kappa_2^Q$, or $\kappa_{11}^{pQ}/\kappa_{11}^{kQ}$ are
    unaffected by global conservation laws.
    However, similar ratios involving cumulants of two non-conserved quantities, such as $\kappa_{11}^{\pi p}/\kappa_{2}^{p}$, do depend on the acceptance.
    This is a useful observation for present~\cite{Adam:2019xmk} and future measurements of the off-diagonal cumulants of net-particle distributions.
    We note that experimental measurements should be performed such that the acceptance parameter $\alpha$ is the same for all hadron species that go into the measurement. Ideally, this entails $p_T$-integrated measurements in a finite rapidity $Y$ acceptance, as opposed to the currently available measurements in a finite $p_T$ range and/or pseudorapidity $\eta$ acceptance~\cite{Adam:2019xmk}.
    
    \item The second order cumulant $\kappa_2^p$ of a non-conserved quantity, such as e.g. net-proton number $p$, in acceptance $\alpha$ represents a linear combination between the grand canonical~($\alpha \to 0$) and canonical~($\alpha \to 1$) limits~[Eq.~\eqref{eq:kappapp}]. Furthermore, the canonical susceptibility~(cumulant) is expressed solely in terms of the matrix of second order grand canonical susceptibilities involving the conserved charges and a non-conserved quantity~[Eq.~\eqref{eq:tildechi}]. In Sec.~\ref{sec:netPnetL} we used this expression to analyze the influence of the various conservation laws on net-proton and net-$\Lambda$ fluctuations in a HRG model.
    We found that, in addition to baryon number conservation, the variances of net-proton and net-$\Lambda$ fluctuations are markedly influenced by net-charge and net-strangeness conservation, respectively. This is a new element compared to prior HRG model studies~\cite{Bzdak:2012an,Braun-Munzinger:2016yjz,Braun-Munzinger:2020jbk} that considered only the effect of baryon number conservation on these quantities.
    
    \item Our studies in the present paper have been focused on a scenario where the total system volume is fixed in all events. Event-by-event fluctuations of the system volume, on the other hand, cannot be avoided completely in heavy-ion collisions and they have their own influence on fluctuations of conserved charges. 
    Different methods exist to address these~\cite{Gorenstein:2011vq,Skokov:2012ds,Braun-Munzinger:2016yjz}.
    In this paper, we have shown in Sec.~\ref{sec:SIQ} that a ratio of strongly intensive measures $\Sigma$ and $\Delta$ involving any two conserved charges is insensitive to both the global charge conservation and volume fluctuations. 
    While the physical interpretation of strongly intensive measures is somewhat less straightforward than that of the traditional cumulants,
    their insensitivity to volume fluctuations makes them useful observables when volume fluctuations are difficult to control.
    The concept of strongly intensive cumulants, introduced in Ref.~\cite{Sangaline:2015bma}, can be used to extend these considerations to higher-order fluctuation measures.
    
    \item The SAM is formulated to study the effects of global charge conservation on cumulants measured in the coordinate space.
    Experimental measurements in heavy-ion collisions, on the other hand, are performed in the momentum space.
    Nevertheless, strong space-momentum correlations at the highest collision energies due to longitudinal Bjorken flow allow to associate measurements in finite rapidity space with spatial subvolumes at the freeze-out stage.
    Furthermore, the fluctuation measures where global conservation factors were found to cancel out can be expected to be robust probes of the grand canonical susceptibilities even in the absence of strong space-momentum correlations.
    
    \item We focused the discussion on the effects of global conservation laws. It is not unfeasible, however, that the exact conservation of charges takes place not only globally, but also in localized spatial regions, as discussed in a number of recent papers~\cite{Castorina:2013mba,Vovchenko:2018fiy,Oliinychenko:2019zfk,Pruneau:2019baa,Vovchenko:2019kes,Braun-Munzinger:2019yxj,Oliinychenko:2020cmr,Altsybeev:2020qnd}. 
    The SAM can be applied in such a scenario if these localized spatial regions
    -- called patches according to the terminology developed in
    Refs.~\cite{Oliinychenko:2019zfk,Oliinychenko:2020cmr} -- are regarded as total volumes $V$
    where the conserved charges are conserved exactly.
    The only requirement is that patches are sufficiently large to contain all physics associated with the correlation length.

\end{itemize}

To conclude, we developed a formalism to quantify the effect of global conservation of multiple conserved charges on fluctuation measurements in heavy-ion collisions.
In particular, this has allowed us, for the first time, to construct fluctuation measures that, to a leading order, are insensitive to effects of global charge conservation.
In the future we plan to apply the concepts developed in this paper to construct a sampler of an \emph{interacting} hadron resonance gas that preserves \emph{local} correlations and fluctuations encoded in the equation of state, and which can be used in state-of-the-art hydrodynamic simulations of heavy-ion collisions.


\acknowledgments
We thank M.~Gazdzicki for a suggestion to look into the strongly intensive quantities.
V.V. was supported by the
Feodor Lynen program of the Alexander von Humboldt
foundation.
This work received support through the U.S. Department of Energy, 
Office of Science, Office of Nuclear Physics, under contract number 
DE-AC02-05CH11231231 and received support within the framework of the
Beam Energy Scan Theory (BEST) Topical Collaboration.
R.P. acknowledges the support by the Stiftung Polytechnische Gesellschaft Frankfurt and the Program of Fundamental Research of the Department of Physics and Astronomy of the National Academy of Sciences of Ukraine and thanks the Frankfurt Institute for Advanced Studies for its hospitality.


\appendix

\section{Evaluation of the higher-order cumulants}
\label{app:highorder}

The evaluation of the fourth and higher-order cumulants of conserved charges in the SAM proceeds by iteratively differentiating the third order cumulants~[Eq.~\eqref{eq:kappa3t}],
\eq{\label{eq:app:kappa3t}
\hat{\tilde{\kappa}}_{i j k}(\hat{t}) & =
\frac{1}{\alpha^2 V^2T^6} \, 
\hat{\tilde{\kappa}}_{ij_1} \, 
\hat{\tilde{\chi}}^{'-1}_{j_1 m_1} \, 
\hat{\tilde{\chi}}^{'}_{m_1 m_2 m_3} \, 
\hat{\tilde{\chi}}^{'-1}_{m_3 m_4} \,  
\hat{\tilde{\kappa}}_{m_4 k} \,
\hat{\tilde{\chi}}^{'-1}_{m_2 j_2} \,
\hat{\tilde{\kappa}}_{j_2j} \nonumber \\
& \quad 
- \frac{1}{\beta^2 V^2T^6}
\hat{\tilde{\kappa}}_{ij_1} \,
\hat{\tilde{\chi}}^{''-1}_{j_1 m_1} \, 
\hat{\tilde{\chi}}^{''}_{m_1 m_2 m_3} \, 
\hat{\tilde{\chi}}^{''-1}_{m_3 m_4} \,  
\hat{\tilde{\kappa}}_{m_4 k} \,
\hat{\tilde{\chi}}^{''-1}_{m_2 j_2} \,
\hat{\tilde{\kappa}}_{j_2j}~,
}
with respect to $\hat{t}$.
For instance the fourth order cumulants are defined as
\eq{\label{eq:app:kappa4tdef}
\hat{\tilde{\kappa}}_{i j k l}(\hat{t}) = \frac{\partial \hat{\tilde{\kappa}}_{ijk} (\hat{t})}{\partial \hat{t}_{l}}~.
}

The two terms in Eq.~\eqref{eq:app:kappa3t} have the structure of a convolution of several factors.
Each factor is either ($i$) a cumulant $\hat{\kappa}$, ($ii$) an inverse second order susceptibility $\hat{\tilde{\chi}}^{'-1}$ or $\hat{\tilde{\chi}}^{''-1}$, or ($iii$) a susceptibility $\hat{\tilde{\chi}}^{'}$ or $\hat{\tilde{\chi}}^{''}$.
For ($i$) and ($iii$) we shall consider $\hat{\kappa}$ and $\hat{\tilde{\chi}}^{'('')}$ to be of an arbitrary order.
The derivative~\eqref{eq:app:kappa4tdef} is computed by applying the product rule to each of the two terms in the r.h.s. of Eq.~\eqref{eq:app:kappa3t}.
There are three kinds of $\hat{t}$-derivatives to compute:
\begin{itemize}
    \item The derivatives of $\hat{\tilde{\kappa}}_{i_1\ldots i_M}$. These simply yield a cumulant of a higher order by definition:
    \eq{\label{eq:app:derkappa}
    \frac{\partial \hat{\tilde{\kappa}}_{i_1\ldots i_M}}{\partial \hat{t}_{i_{M+1}}} \equiv \hat{\tilde{\kappa}}_{i_1\ldots i_{M+1}}~.
    }
    \item The derivatives of $\hat{\tilde{\chi}}_2^{'-1}$ and $\hat{\tilde{\chi}}_2^{''-1}$. 
    These have already been computed in Sec.~\ref{sec:kappa3}. 
    The corresponding expressions are given by Eqs.~\eqref{eq:derinvchi1} and \eqref{eq:derinvchi2} which we rewrite here for completeness:
    \eq{\label{eq:app:derinvchi1}
    \frac{ \partial \hat{\tilde{\chi}}^{'-1}_{j_1j_2} }{\partial \hat{t}_k} 
    & = -\frac{1}{\alpha V T^3} \,
    \hat{\tilde{\chi}}^{'-1}_{j_1 m_1} \, 
    \hat{\tilde{\chi}}_{m_1 m_2 m_3}^{'} \, 
    \hat{\tilde{\chi}}^{'-1}_{m_3 m_4} \,  
    \hat{\tilde{\kappa}}_{m_4 k} \,
    \hat{\tilde{\chi}}^{'-1}_{m_2 j_2}~,\\
    \label{eq:app:derinvchi2}
    \frac{ \partial \hat{\tilde{\chi}}^{''-1}_{j_1j_2} }{\partial \hat{t}_k} & = \frac{1}{\beta V T^3} \,
    \hat{\tilde{\chi}}^{''-1}_{j_1 m_1} \, 
    \hat{\tilde{\chi}}_{m_1 m_2 m_3}^{''} \, 
    \hat{\tilde{\chi}}^{''-1}_{m_3 m_4} \,  
    \hat{\tilde{\kappa}}_{m_4 k} \,
    \hat{\tilde{\chi}}^{''-1}_{m_2 j_2}~.
    }
    \item The derivatives of $\hat{\tilde{\chi}}^{'}_{i_1\ldots i_M}$ and $\hat{\tilde{\chi}}^{''}_{i_1\ldots i_M}$.
    These are evaluated by applying the chain rule. The result is:
    \eq{\label{eq:app:derchi1}
    \frac{\partial \hat{\tilde{\chi}}_{i_1\ldots i_M}^{'}}{\partial \hat{t}_{i_{M+1}}} & = \frac{1}{\alpha V T^3} \hat{\tilde{\chi}}_{i_1\ldots i_M b_1}^{'} \hat{\chi}^{'-1}_{b_1b_2} \hat{\tilde{\kappa}}_{b_2 i_{M+1}}~,\\
    \label{eq:app:derchi2}
    \frac{\partial \hat{\tilde{\chi}}_{i_1\ldots i_M}^{''}}{\partial \hat{t}_{i_{M+1}}} &= -\frac{1}{\beta V T^3} \hat{\tilde{\chi}}_{i_1\ldots i_M b_1}^{''} \hat{\tilde{\tilde{\chi}}}^{''-1}_{b_1b_2} \hat{\tilde{\kappa}}_{b_2 i_{M+1}}~.
    }
\end{itemize}

As follows from the rules~\eqref{eq:app:derkappa}-\eqref{eq:app:derchi2}, a $\hat{t}$-derivative of a term which comprises a convolution of an arbitrary number of $\hat{\tilde{\kappa}}$, $\hat{\tilde{\chi}}^{-1}_2$, and $\hat{\tilde{\chi}}$ elements yields a sum of terms, each again being a certain convolution of $\hat{\tilde{\kappa}}$, $\hat{\tilde{\chi}}^{-1}_2$, and $\hat{\tilde{\chi}}$ elements.
Therefore, the rules~\eqref{eq:app:derkappa}-\eqref{eq:app:derchi2} are sufficient to iteratively compute $\hat{t}$-derivatives of $\hat{\tilde{\kappa}}_{ijk} (\hat{t})$ up to arbitrary high order.
Further simplifications can be achieved by observing that all the $\hat{\kappa}$, $\hat{\tilde{\chi}}^{-1}_2$, and $\hat{\tilde{\chi}}$ tensors are symmetric with respect to any permutation of their indices.

We have implemented the above rules within the \texttt{Mathematica} package to compute the higher order cumulants $\hat{\tilde{\kappa}}$.
Equations~\eqref{eq:kappa4final}, \eqref{eq:kappa5final}, and \eqref{eq:kappa6final} in the main text depict the final results for the fourth, fifth, and sixth order cumulants, respectively, all evaluated at $\hat{t} = 0$.

\section{Deriving QCD cumulants from the general expressions}
\label{app:illustr}

Here we illustrate how to derive cumulants of baryon number $B$ and electric charge $Q$ from the general expressions given in Eqs.~\eqref{eq:kappa1final}-\eqref{eq:kappa6final}.
This is useful to illustrate the SAM notation entering Eqs.~\eqref{eq:kappa1final}-\eqref{eq:kappa6final} which differs from the commonly employed notations in the QCD literature.

Let us have two conserved charges: baryon number $B$ and electric charge $Q$.
In this case the vector of conserved charges reads $\hat{Q} = (B,Q)$ and the corresponding vector of chemical potentials is $\hat{\mu} = (\mu_B,\mu_Q)$.
Following Eq.~\eqref{eq:cumudef}, a grand canonical susceptibility of order $M$ reads
\eq{\label{eq:suscdefBQ}
\hat{\chi}_{i_1\ldots i_M}
~=~\frac{\partial^{M}(p/T^4)}{\partial(\mu_{i_1}/T) \, \dots \, \partial(\mu_{i_M}/T)}~, 
\qquad i_1\ldots i_M \in 1,2.
}

In the commonly adopted QCD notation, the same grand canonical susceptibility reads
\eq{\label{eq:suscdefBQ2}
\chi^{BQ}_{lm} = \frac{\partial^{l+m}(p/T^4)}{\partial(\mu_{B}/T)^l \, \partial(\mu_{Q}/T)^m}~,
\qquad l+m = M.
}

Definitions~\eqref{eq:suscdefBQ} and \eqref{eq:suscdefBQ2} are equivalent, provided that exactly $l$ of the indices $(i_1,\ldots,i_M)$ are equal to unity while the remaining $m$ indices are equal to two.
Note that the susceptibility in Eq.~\eqref{eq:suscdefBQ} is symmetric with respect to any permutation of its indices, therefore, it is irrelevant how exactly the indices are distributed in the vector $(i_1,\ldots,i_M)$.
The same discussion of the two different notations as for the susceptibilities also applies to cumulants.

The general results in Eqs.~\eqref{eq:kappa1final}-\eqref{eq:kappa6final} for the cumulants of conserved quantities in a subvolume use the notation~\eqref{eq:suscdefBQ}.
Let us illustrate how to derive the results for cumulants in a more familiar notation~\eqref{eq:suscdefBQ2} on an example of fourth order cumulants.
The general expression~\eqref{eq:kappa4final} for a fourth order cumulant of conserved charges $B$ and $Q$ reads
\eq{\label{eq:kappa4BQ}
\hat{\kappa}_{i_1 i_2 i_3 i_4}[B^1,Q^1] & = \alpha V T^3 \,  \beta \left[
\, (1-3\alpha \beta) \, \hat{\chi}_{i_1i_2i_3i_4} - \frac{\alpha \beta}{2!\, 2!\, 2!} \sum_{\sigma \in S_4} \hat{\chi}^{-1}_{b_1b_2} \, \hat{\chi}_{\per{1}\per{2}b_1} \,  \hat{\chi}_{\per{3}\per{4}b_2} \, \right].
}
The repeated indices $b_1$ and $b_2$ imply summation from 1 to 2.
The sum $\sum_{\sigma \in S_4}$ runs over the 4!=24 permutations of a set $(1,2,3,4)$. 
For instance, the first element of the sum corresponds to $\sigma = (1,2,3,4)$, so that $i_{\sigma_1} = i_1$, $i_{\sigma_2} = i_2$, $i_{\sigma_3} = i_3$, and $i_{\sigma_4} = i_4$.
For the second permutation one has $\sigma = (1,2,4,3)$, so that $i_{\sigma_1} = i_1$, $i_{\sigma_2} = i_2$, $i_{\sigma_3} = i_4$, and $i_{\sigma_4} = i_3$. And so on for all remaining permutations.

Consider now the diagonal fourth order cumulant of baryon number $\kappa_4[B^1] \equiv \kappa_{4}^{B} \equiv \kappa_{40}^{BQ}$~(here we omit the superscript for $B^1$ and $Q^1$ for clarity of notation).
In this case one has $i_1 = i_2 = i_3 = i_4 = 1$ in Eq.~\eqref{eq:kappa4BQ}, thus 
\eq{\label{eq:kappa4BQ1}
\kappa_4[B^1] & \equiv \hat{\kappa}_{1111}[B^1,Q^1] \nonumber \\
& = \alpha V T^3 \,  \beta \left[
\, (1-3\alpha \beta) \, \hat{\chi}_{1111} - \frac{\alpha \beta}{2!\, 2!\, 2!} \sum_{\sigma \in S_4} \hat{\chi}^{-1}_{b_1b_2} \, \hat{\chi}_{11b_1} \,  \hat{\chi}_{11b_2} \, \right].
}

Each of the $\sigma \in S_4$ permutations yields the same element in Eq.~\eqref{eq:kappa4BQ1}, therefore, one has
\eq{\label{eq:kappa4BQ2}
\kappa_4[B^1] & = \alpha V T^3 \,  \beta \left[
\, (1-3\alpha \beta) \, \hat{\chi}_{1111} - 3 \alpha \beta \, \hat{\chi}^{-1}_{b_1b_2} \, \hat{\chi}_{11b_1} \,  \hat{\chi}_{11b_2} \, \right].
}

Given the correspondence between Eqs.~\eqref{eq:suscdefBQ} and \eqref{eq:suscdefBQ2}, we have 
$\hat{\chi}_{1111} \equiv \chi^{BQ}_{40} \equiv \chi^B_4$,
$\hat{\chi}_{111} \equiv \chi^{BQ}_{30} \equiv \chi^B_3$, 
and $\hat{\chi}_{112} \equiv \chi^{BQ}_{21}$.
The direct, $\hat{\chi}_{b_1 b_2}$ and the inverse, $\hat{\chi}^{-1}_{b_1b_2}$, matrices of the second order susceptibilities read
\eq{\label{eq:chi2BQ}
\hat{\chi}_2  = 
\begin{pmatrix}
 \chi_2^B & \chi_{11}^{BQ} \\
 \chi_{11}^{BQ} & \chi_2^Q
\end{pmatrix},~~~~~~
\hat{\chi}_2^{-1}  = \frac{1}{\det[\hat{\chi}_2]}
\begin{pmatrix}
 \chi_2^Q & -\chi_{11}^{BQ} \\
 -\chi_{11}^{BQ} & \chi_2^B
\end{pmatrix},~~~~~~
\det[\hat{\chi}_2] & = \chi_2^B \, \chi_2^Q - (\chi_{11}^{BQ})^2.
}

Equation~\eqref{eq:kappa4BQ2} transforms to
\eq{\label{eq:kappa4BQ3}
\kappa_4[B^1] = \alpha VT^3 \, \beta \, \left[ 
 \, (1-3\alpha \beta) \, \chi_4^B
- 3 \, \alpha \, \beta \, \frac{ (\chi_3^B)^2 \chi_2^Q - 2 \chi_{21}^{BQ} \chi_{11}^{BQ} \chi_3^B + (\chi_{21}^{BQ})^2 \chi_2^B }{\chi_2^B \chi_2^Q - (\chi_{11}^{BQ})^2}
\right].
}

Explicit relations listed in Sec.~\ref{sec:BQS} for various cumulants of the QCD conserves charges are obtained from Eqs.~\eqref{eq:kappa1final}-\eqref{eq:kappa6final} in the same fashion as for $\kappa_4[B^1]$ shown here.

A \texttt{Mathematica} notebook to express any cumulant up to sixth order within the QCD notation is available via~\cite{SAMgithub}.

\bibliographystyle{JHEP}
\bibliography{subensembleBQS}

\end{document}